\documentclass[pre,aps,twocolumn]{revtex4-1}
\usepackage{amsmath,amssymb}
\usepackage{graphicx}
\usepackage{dcolumn}
\usepackage{bm}
\usepackage{verbatim}
\usepackage{color}

\newlength{\imgwidth}
\newlength\subfigheight
\newlength\subfigtextheight

\begin{document}

\title{Phase behavior and structure of colloidal bowl-shaped particles: simulations}

\author{Matthieu Marechal}
\author{Marjolein Dijkstra}
\email{M.Dijkstra1@uu.nl}

 \affiliation{Soft Condensed Matter, Debye Institute for NanoMaterials Science, Utrecht University,
  Princetonplein 5, 3584 CC Utrecht, The Netherlands}
\date{\today}

\begin{abstract}
We study the phase behavior of bowl-shaped particles using
computer simulations. 
These particles were found experimentally to form 
a meta-stable worm-like fluid phase in which the
bowl-shaped particles have a strong tendency to stack on top of
each other [M.Marechal \emph{et al}, Nano Letters \textbf{10}, 1907 (2010)]. In this work, we show that 
the transition from the low-density fluid to the worm-like phase 
has an interesting effect on the equation of state.
The simulation results also show that the worm-like fluid phase transforms spontaneously into a columnar phase for bowls that
are sufficiently deep.
Furthermore, we describe the phase behavior as obtained from
free energy calculations employing Monte Carlo simulations. 
The columnar phase is stable for bowl shapes ranging from infinitely thin bowls 
to surprisingly shallow bowls. Aside from a large region of stability for the columnar phase, 
the phase diagram features four novel crystal phases and a region where the stable fluid contains worm-like stacks.
\end{abstract}

\maketitle

\section{Introduction}

The concept of a mesogenic particle in the form of a bowl is
relatively old in the molecular liquid crystal community. Such
molecules are expected to form a columnar phase, which can be
ferroelectric, i.e., a phase with a net electric dipole moment,
when the particles possess a permanent dipole moment.
Ferroelectric phases have potential applications for optical and
electronic devices.
In fact, crystalline (as opposed to liquid crystalline) ferroelectrics are already applied in sensors, electromechanical devices
and non-volatile memory~\cite{FerroApp}. A columnar ferroelectric phase may have the advantage over a crystal, that 
grain boundaries and other defects anneal out faster due to the partially fluid nature of the columnar phase.
In reality, columnar phases of conventional disc-like particles often
exhibit many defects, as flat thin discs can  diffuse out of a column
and columns can split up. The presence of these defects limits their
potential use for industrial applications~\cite{simulation-bowls}.
Less  defects are expected in a columnar phase of
bowl-shaped mesogens, where particles are supposed to be more confined
in the lateral directions. A whole variety of bowl-like molecules
have already been synthesized and investigated
experimentally~\cite{Sawamura2002,simpson2004,xu1993rbl,malthete1987icc}. In addition, buckybowlic molecules, \emph{i.e.} fragments of $C_{60}$ whose dangling bonds have been saturated with hydrogen atoms, have been shown to crystallize
in a columnar fashion~\cite{Rabideau1996,Forkey1997,Matsuo2004,Sakurai2005,Kawase2006}. 
However, the number of theoretical studies is very limited as it
is difficult to model the complicated particle shape in theory and
simulations. In a recent simulation study, the
attractive-repulsive Gay-Berne potential generalized to bowl-shaped particles has been used to investigate the stacking of
bowl-like mesogens as a function of temperature~\cite{simulation-bowls}.
The authors reported a nematic phase and a columnar phase. 
This columnar phase did not exhibit overall ferroelectric order, although polar regions were found.
In another very recent simulation study~\cite{Cinacchi2010} of hard contact lenses
(infinitely thin, shallow bowls), a new type of fluid phase was found in which the particles cluster on a spherical surface for bowls which are not too shallow.
No columnar phase was found since the focus was on rather shallow bowls at a relatively low densities. 

Recently, a procedure has been developed to
synthesize bowl-shaped colloidal particles~\cite{Carmen}. This
method starts with the preparation of highly uniform 
oil-in-water emulsion droplets. Subsequently, the droplets
were used as templates around which a solid shell with tunable thickness is grown. In the next step of the synthesis, the
oil in the droplets is dissolved and finally, during
drying, the shells collapse into hemispherical
double-walled bowls. 
In addition to these larger, more easily imaged colloids, a whole variety of bowl-shaped nanoparticles and smaller colloids have been synthesized and characterized%
~\cite{Charnay2003,Wang2004,Liu2005,Jagadeesan2008,Love2002,Hosein2007}, and possible applications of these systems have been put forward.
We also note
that recently hemispherical particles were synthesized at an
air-solution interface~\cite{higuchi} and on a
substrate~\cite{Xia}. These hemispherical particles are intended
to be used as microlense arrays, but they can also serve as a new
type of shape-anisotropic colloidal particle.

\begin{figure}[t]

\includegraphics[width=0.45\textwidth]{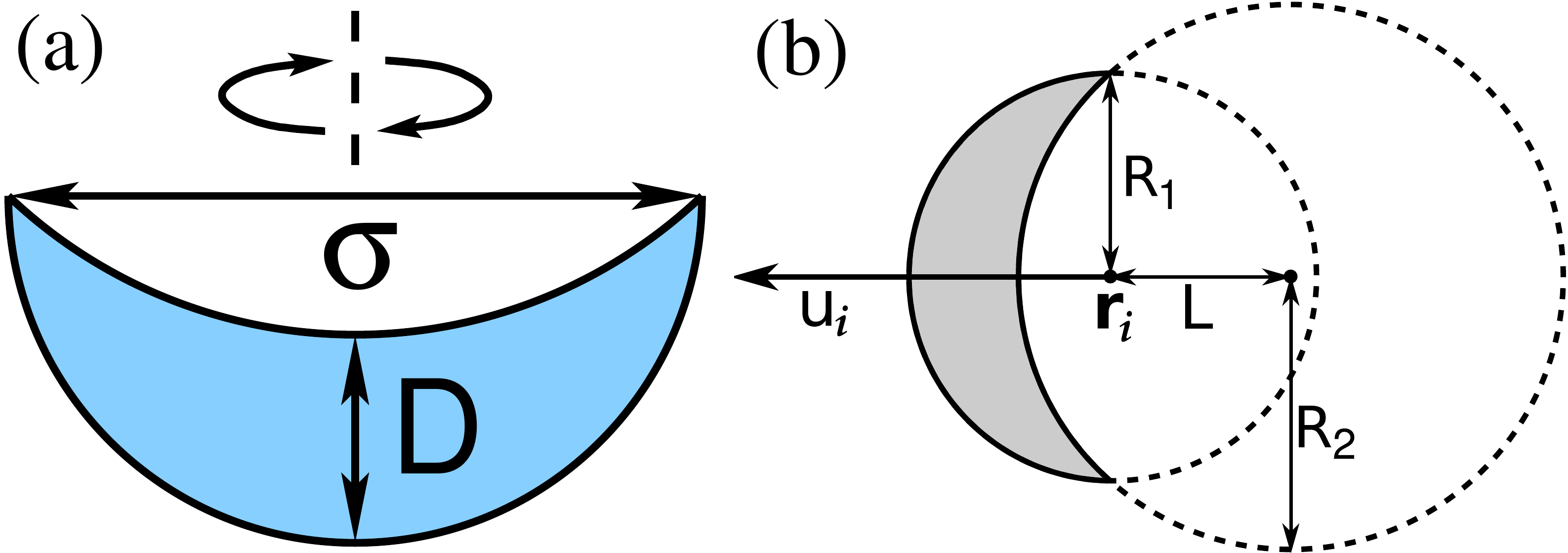}

\caption{ 
(a) The theoretical model of the
colloidal bowl is the solid of revolution of a crescent around the
axis indicated by the dashed line.  The thickness of the
double-walled bowl is denoted by $D$ and the diameter of the bowl
by $\sigma$.
(b) The bowls are defined using two spheres of radii $R_1$ and $R_2$, that are a distance of $L$ apart. 
The direction vector, $\mathbf{u}_i$ and the reference point of the particle, $\mathbf{r}_i$,
(the dot in the center of the smaller sphere) are indicated. 
\label{fig:particles}}
\end{figure}

In our simulations, we model the particles as the solid of
revolution of a crescent
(see
 Fig.~\ref{fig:particles}a).
The diameter $\sigma$ of the particle and the thickness $D$ are
defined as indicated in Fig.~\ref{fig:particles}a. We define the
shape parameter of the bowls by a reduced thickness $D/\sigma$,
such that the model reduces to infinitely thin hemispherical
surfaces for $D/\sigma=0$ and to solid hemispheres for
$D/\sigma=0.5$. The advantages of this simple  model is that it
interpolates continuously between an infinitely thin bowl and a
hemispherical solid particle (the two colloidal model systems,
which we discussed above), and that we can derive an algorithm
that tests for overlaps between pairs of bowls, which is a
prerequisite for Monte Carlo simulations of hard-core systems.

In a recent combined experimental and simulation study (for which we performed the simulations), the phase behavior of repulsive bowl-shaped colloids was investigated~\cite{Marechal2010bowls}.  
The colloids were shown to form a worm-like fluid phase, in which the particles form long curved stacks running in random directions.
By comparing the distribution of stack lengths, the simulation model was shown to describe the colloidal particles well. 
No evidence of columnar ordering was found in the experiments and in simulations of bowls with corresponding deepness,
which was explained by the glassy behavior of the particles preventing rearrangements.
The phase behavior of the model particles is expected to also describe other repulsive bowl shaped particles well, provided that the dimensions of the simulation
particle are chosen such that the diameter of a stack and the inter-particle distance in the stack are the same as for the particles to be modeled. 

In this work, we expand the simulation results on the hard bowl-shaped particles.
First, we elaborate on the model for the collapsed shells; the overlap algorithm is described in the appendix.
Also, the (free energy) methods are explained in more detail than in Ref.~\cite{Marechal2010bowls}.
In the results section, we study the properties of the isotropic phase. We investigate the nature and the location of the transition between the homogeneous
fluid phase
and the fluid phase that contains the worm-like stacks. 
Furthermore, we show the packing diagram and the phase diagram with a tentative homogeneous--to--worm-like fluid transition line. In the last section we summarize and discuss the results.

\section{Methods}

\subsection{Model}

We describe the model that we use to represent the 
bowls in more detail. Consider a sphere with a radius $R_1$ at the
origin and a second sphere with radius $R_2>R_1$ at position
$-L\mathbf{u}_i$, where $\mathbf{u}_i$ is the unit vector denoting
the orientation of the bowl and $L>0$. The bowl is represented by
that part of the sphere with radius $R_1$ that has no overlap with
the larger sphere, see Fig.~\ref{fig:particles}b. We have chosen the
values for $L$ and $R_2$ such that the bowls are hemispherical (see appendix for explicit expressions
for $L$ and $R_2$).
We define the thickness of
the bowls  by $D=L-(R_2-R_1)$, such that  the model reduces to the
surface of a hemisphere for $D=0$ and  to a solid hemisphere for
$D=R_1$. The volume of the particle is $\frac{\pi}{4}\, D\,
(\sigma^2 - D\sigma + \frac{2}{3} D^2)$, where $\sigma\equiv 2R_1$
is our unit of length. The algorithm to determine overlap between
our bowls is described in the appendix.

\subsection{Fluid phase}

We employ standard $NPT$ MC simulations to obtain the equation of
state (EOS) for the fluid phase. In addition, we obtain the
compressibility by measuring the fluctuations in the volume:
\begin{equation}
\frac{\langle V^2\rangle - \langle V\rangle^2}{\langle V\rangle}=\frac{k_B T}{\rho} \, \left(\frac{\partial\rho}{\partial P}\right)_T,
\end{equation}
where $\rho=N/V$ is the number density and the derivative of the pressure is taken at constant temperature is denoted by the subscript $T$.
We determine the free energy at density $\rho_1$  by integrating
the EOS from  reference density $\rho_0$ to $\rho_1$:
\begin{equation}
\frac{F(\rho_1)}{N}=
\mu(\rho_0)-\frac{P(\rho_0)}{\rho_0} + \int_{\rho_0}^{\rho_1} \frac{P(\rho)}{\rho^2} d \rho\label{eqn:widom}
\end{equation}
where the chemical potential $\mu(\rho_0)$ is determined using the
 Widom particle insertion method~\cite{widom}, and $P(\rho_0)$ is determined by a local fit to the EOS.

To investigate the structure of the fluid phase, we measure the
positional correlation function~\cite{Veerman_Frenkel},
\begin{equation}
g_c(z)=\frac{1}{N \rho A_\text{col}} \langle \sum_{i=1}^N
\sum_{j=1}^{N_\text{col}(i)} \delta(\mathbf{r}_{ij} \cdot
\mathbf{u}_i-z) \rangle\label{eqn:gcz},
\end{equation}
where the sum over $j$ runs over  $N_\text{col}(i)$ particles in a
column of radius $\sigma/2$ with orientation $u_i$ centered around
particle $i$, and where the area of the column is denoted by
$A_\text{col}=\pi \sigma^2/4$. At sufficiently high pressure the
particles stack on top of each other to form disordered  worm-like
piles which resemble the stacks observed in the experiments~\cite{Marechal2010bowls}.
As the stacks have a strong
tendency to buckle, we cannot use $g_c(z)$ to determine the length
of the stacks. We therefore determine the stack size distribution
using a cluster criterion. Particle $i$ and $j$ belong to the same
cluster if
\begin{eqnarray}
 |\mathbf{r}_{ij} + (\zeta D/2 + \sigma/4) (\mathbf{u}_j-\mathbf{u}_i)| & < & \sigma/2 \quad\text{and} \nonumber\\
\mathbf{u}_i \cdot \mathbf{u}_j & > & 0, \label{eqn:cluster_crit}
\end{eqnarray}
and where the first condition has to be satisfied for $\zeta=-1$,
$0$ or $1$ and $\mathbf{r}_{ij}=\mathbf{r}_j-\mathbf{r}_i$, with $\mathbf{r}_i$ denoting the center of the
sphere with radius $R_1$ of particle $i$, see Fig.~\ref{fig:particles}b.
If both conditions are satisfied, particle $j$ is just above
($\zeta=1$) or below ($\zeta=-1$) particle $i$ in the stack, or,
when the stack is curved, particle $j$ can be next to particle $i$
($\zeta=0$). We now define the cluster distribution as the
fraction of particles that belongs to a cluster of size $n$:
$\mathcal{P}_\text{stack}(n)\equiv n N_n/N$, where $N_n$ is the number of
clusters of size $n$. We checked that the cluster size
distribution does not depend sensitively to the choice of
parameters in Eq.~(\ref{eqn:cluster_crit}).


\subsection{Columnar phases}

We also perform  $NPT$ Monte Carlo simulations of the columnar phase using a
rectangular simulation box with varying box lengths in order to
relax the inter-particle distance in the $z$ direction, along the
columns, independently from the lattice constant in the
horizontal direction. The difference between the free energy of
the columnar phase at a certain density and the free energy of the
fluid phase at a lower density is determined using a thermodynamic
integration technique~\cite{Bates_Frenkel}. We apply a potential
which couples a particle to its column:
\begin{equation}
\Phi_\mathrm{hex}(\mathbf{r}^N,\lambda)=\lambda\sum_{i=1}^N\cos(2 \pi N_x x_i/L_x)\sin(\pi N_y y_i/L_y), \label{eqn:col_pot}
\end{equation}
where $x_i$ and $y_i$ are the $x$ and $y$ components respectively of $\mathbf{r}_i$, $N_\alpha$ is the number of columns in the $\alpha$ direction and $L_\alpha$ is the size of the box in the $\alpha$ direction.
In our simulations, we calculate Eq.~(\ref{eqn:col_pot}) while fixing the center of mass. 
To do so efficiently, we first calculate all four combinations
\begin{equation}
\lambda\sum_{i=1}^N \mathrm{trig1}(2 \pi N_x x_i/L_x) \mathrm{trig2}(\pi N_y y_i/L_y)\label{eqn:col_gen_trigs}
\end{equation} for $\mathrm{trig1}=\cos,\sin$ and $\mathrm{trig2}=\cos,\sin$. The change in
these four expressions upon displacement of a single particle while keeping the center of mass fixed can be expressed in terms of single particle properties and the previous values of the expressions by using some basic trigonometry.
In this way, $\Phi_\mathrm{hex}(\mathbf{r}^N,\lambda)$, which is
Eq.~\ref{eqn:col_gen_trigs} for $\mathrm{trig1}=\cos$ and $\mathrm{trig2}=\sin$, can be calculated without 
performing the full summation over all particles in Eq.~(\ref{eqn:col_gen_trigs}) every time we displace a particle. 
Unfortunately, this calculation requires the evaluation of many more trigonometric functions than the simple expression (\ref{eqn:col_pot}),
but the extra computation time is negligible compared to the overlap check.

In addition to this positional potential, we also constrain the direction of the particle,
using the potential
\begin{equation}
\Phi_\text{ang}(\mathbf{u}^N,\lambda)=\lambda' \sum_{i=1}^N u_{i,z},\label{eqn:col_pot_ang}
\end{equation}
where we used $\lambda'=0.1\lambda$ and where $u_{i,z}$ is the $z$ component of $\mathbf{u}_i$.
The thermodynamic integration path from the columnar phase to the fluid is as follows: We start from the columnar phase at a certain density $\rho_2$. Subsequently, we slowly turn on the two potentials, \emph{i.e.} we increase $\lambda$ from 0 to $\lambda_\mathrm{max}$.
Next, we integrate the equation of state to go from $\rho_2$ to $\rho_1$, while keeping $\lambda=\lambda_\text{max}$ fixed. During this step
the columnar phase will only be stable below the coexistence density, if $\lambda_\text{max}$ is sufficiently high.
We find that $\lambda_\text{max}=20k_BT$ suffices to guarantee stability of the columnar phase.
Finally, fixing the density $\rho_1$, we gradually turn off the potentials, while integrating over
$\lambda$ from $\lambda_\text{max}$ to 0. During this last step, the columnar phase melts continuously, provided that the density $\rho_1$ is low enough and that
$\lambda$ is high enough to prevent melting during the density integration step.
The resulting free energy difference between the columnar phase and fluid phase is given by
\begin{multline}
F_\text{col}(\rho_2)-F_\text{fluid}(\rho_1)=\\\int_0^{\lambda_\text{max}} \big\langle \Phi_\text{hex}(\mathbf{r}^N,\lambda)/\lambda+\Phi_\text{ang}(\mathbf{u}^N,\lambda)/\lambda\big\rangle\big|_{\rho=\rho_1}+\\
\int_{\rho_1}^{\rho_2} d\rho\left.\frac{NP(\rho)}{\rho^2}\right|_{\lambda=\lambda_\text{max}} \\
-\int_0^{\lambda_\text{max}} \big\langle \Phi_\text{hex}(\mathbf{r}^N,\lambda)/\lambda+\Phi_\text{ang}(\mathbf{u}^N,\lambda)/\lambda\big\rangle\big|_{\rho=\rho_2}
\end{multline}
The positional potential~(\ref{eqn:col_pot}) is designed to stabilize a hexagonal array of columns, but, strictly speaking,
it does not have the hexagonal symmetry of the columnar phase,
since it is not invariant under a 60 degrees rotation of the whole system around a lattice position.
However, we find that replacing Eq.~(\ref{eqn:col_pot}) by a positional potential that does have this symmetry, does not
have a significant effect on the free energy difference.

A second type of columnar phase can be constructed by flipping half of the bowls.
In this way we obtain alternating vertical sheets (\emph{i.e.} rows of columns) of bowls that point upwards and sheets of bowls that point downwards, we will
refer to this phase as the inverted columnar phase.
We calculate the free energy of this phase using the method described above, with the modification that the angular potential now reads,
\begin{equation}
\Phi_\text{ang}(\mathbf{u}^N,\lambda)=\lambda' \sum_i u_{i,z}^2. \label{eqn:col_pot_ang2}
\end{equation}
This potential could also have been used for the non-inverted columnar phase, and we have found that the result of the free energy integration for the columnar phase
is the same whether we use Eq.~(\ref{eqn:col_pot_ang2}) or Eq.~(\ref{eqn:col_pot_ang}).

\subsection{Crystals}
\label{sec:cryst}

\subsubsection{Packing}

As the crystal phases of the bowls are  not known a priori, we developed a novel pressure annealing method
to obtain the possible crystal phases~\cite{PhysRevLettSSS}, which we named after the thermal annealing technique commonly used to find energy minima.
Fully variable box shape $NPT$ simulations were performed
on system of only 2-6 particles. By construction, the final configuration of such a simulation is a crystal, where the unit cell is the simulation box.
One cycle of such a simulation consists of the following steps:
We start at a pressure of $10 k_B T/\sigma^3$. Subsequently, we run a series of simulations, where the pressure increases by a factor of ten each run: $P\sigma^3/k_B T=10,100,\ldots, 10^6$. At the highest pressure ($10^6 k_B T/\sigma^3$) we measure the density and angular order parameters, $S_1\equiv\|\langle \mathbf{u}_i\rangle\|$ and $S_2\equiv \lambda_2$, where $\lambda_2$ is the highest
eigenvalue of the matrix whose components are $Q_{\alpha_\beta}=\frac{3}{2} \langle u_{i\alpha} u_{i\beta} \rangle - \frac{1}{2}\delta_{\alpha\beta}$, where $\alpha,\beta=x,y,z$.
We store the density if it is the highest density found so far for these values of $S_1$ and $S_2$. We ran 1000 of such cycles for each aspect ratio, which is enough
to visit each crystal phase multiple times.
After completing the simulations, we tried to determine the lattice parameters of the resulting crystal by hand. Although this last step
is not necessary, it is convenient to have analytical expressions for the lattice vectors and the density. The pressure annealing runs were performed for $D/\sigma=0.1,0.15,\ldots,0.5$. For many of the crystals, we were not able to find analytical expressions for the lattice parameters. For these crystals, 
we obtain the densities of the close packed crystals for intermittent values of $D$ by averaging the density in single simulation runs at a pressure of $10^6 k_BT/\sigma^3$. 
The initial configurations for the value of $D$ of interest were obtained from the final
configurations of the pressure annealing simulations for another value of $D$ by one of the following
two methods, depending on whether we needed to decrease or increase $D$:
When decreasing $L$ no overlaps are created so the final configuration of the simulation for the previous value of $L$ can be used as initial configuration. On the other hand, increasing $L$ results in an overlap, which is removed by scaling the system uniformly. Subsequently, the pressure is stepwise increased from 1000$k_BT/\sigma^3$ to $10^6 k_BT/\sigma^3$, by multiplying by 10 each step.

\subsubsection{Free energies}

We calculate the free energy of the various
crystal phases by  thermodynamic integration using the Einstein
crystal as a reference state \cite{FrenkelSmit}.  The Einstein
integration scheme that we employ here is similar to the one that was used to calculate the free energies of crystals of dumbbells in Ref.~\cite{dumbbell_article}.
We briefly sketch the integration scheme here and discuss the modifications that we applied.
We couple both the positions and the direction of the particles with a coupling strength $\lambda$, such that for 
$\lambda\rightarrow \infty$, the particles are in a perfect crystalline configuration. First, we integrate $\partial F/\partial\lambda$ over $\lambda$ from zero to a large but finite value
for $\lambda$. Subsequently we replace the hard-core particle--particle interaction potential by a soft interaction, where we can tune the softness of the potential
by the interaction strength $\gamma$. We integrate over $\partial F/\partial \gamma$ from a system with essentially hard-core interaction (high $\gamma=\gamma_\text{max}$), to an ideal Einstein crystal 
($\gamma=0$).  Some minor alterations to the scheme of Ref.~\cite{dumbbell_article} were introduced, which were necessary, because of the different shape of the particle.
For
 the coupling of the
orientation of bowl $i$, \emph{i.e.}, $\mathbf{u}_i$, to an aligning
field, we have to take into account 
that the bowls have no up down symmetry, while the dumbbells are symmetric under $\mathbf{u}_i\rightarrow -\mathbf{u}_i$. The potential energy function that 
achieves the usual harmonic coupling of the particles to their lattice positions, as well as the new angular coupling, reads: 
\begin{multline} \beta U({\bf r}^N,{\bf u}^N;\lambda) =\\
\lambda
\sum_{i=1}^{N} ({\bf r}_i-{\bf r}_{0,i})^2/\sigma^2
 + \sum_{i=1}^N \lambda
(1-\cos(\theta_{i0})) 
, \label{eqn:einstein}
\end{multline}
where ${\bf r}_i$ and ${\bf u}_i$ denote, respectively, the center-of-mass
position and orientation of bowl $i$ and ${\bf r}_{0,i}$ the
lattice site of particle $i$,
 $\theta_{i0}$ is the angle between $\mathbf{u}_i$ and the
ideal tilt vector of particle $i$, and $\beta=1/k_BT$. 
The Helmholtz free energy~\cite{dumbbell_article} of the noninteracting Einstein crystal is modified accordingly, but the only modification is
the integral over the angular coordinates:
\begin{equation}
J(\lambda)=
\int_{-1}^1 e^{\lambda (x-1)} dx=\frac{1-e^{-2\lambda}}{\lambda}.
\end{equation}

Although the shape of the bowls is more complex than that of the dumbbell, we can still use a rather simple form 
for the pairwise soft potential interaction:
\begin{equation}
\beta U_{\mathrm{soft}}({\bf r}^N,{\bf
u}^N;\gamma)=\sum_{i<j} \beta \varphi({\bf
r}_{i}-{\bf r}_{j},\mathbf{u}_i,\mathbf{u}_j,\gamma)
\end{equation} with
\begin{multline}
\beta\varphi({\bf r}_{j}-{\bf r}_{i},\mathbf{u}_i,\mathbf{u}_j,\gamma) =\\
\left\{\begin{array}{cc}  \gamma
(1-A (r_{ij}'/\sigma_\text{max})^2) & \text{if $i$ and $j$ overlap}\\
0 & \mathrm{otherwise}\end{array} \right. , \label{eqn:softpot_part}
\end{multline}
where $r_{ij}'\equiv | \mathbf{r}_j-\mathbf{r}_i+\frac{\sigma-D}{2}(\mathbf{u}_i-\mathbf{u}_j)|$ \emph{i.e.} the distance between the ``centers'' of bowl $i$ and bowl $j$,
$\sigma_\text{max}$ is the maximal $r_{ij}'$ for which the particles overlap: $\sigma_\text{max}^2=\sigma^2+(\sigma-D)^2$,
$A$ is an adjustable parameter that is kept fixed during the simulation
at a value $A=0.5$, and $\gamma$ is the integration parameter.
It was shown in Ref.~\cite{Fortini_soft_pot} that in order to minimize the
error and maximize the efficiency of the free energy calculation,
the potential must decrease as a function of $r$ and must exhibit a discontinuity
at $r$ such that both
the amount of overlap and the number of overlaps decrease upon
increasing $\gamma$.
Here, we have chosen this particular form of
the potential because it can be evaluated very efficiently in a
simulation, although it does not describe the amount of overlap between bowls $i$ and $j$ very accurately.
We checked that adding a term that tries to describe the angular behavior of the amount of overlap does not significantly change our results of the free energy calculations.
Also, we checked that by employing the usual Einstein integration method (\emph{i.e.} only hard-core interactions) at a relatively low density we obtained the same result as by using the method of Fortini {\em et al.}\cite{Fortini_soft_pot}. Finally, we set the maximum interaction strength $\gamma_\text{max}$ to 200.

We perform variable box shape NPT simulations~\cite{Parrinello} to obtain the equation of state
for varying $D$. 
In these simulations not only the edge length changes, but also the angles between the edges are allowed to change.
We employ the averaged configurations in the Einstein crystal thermodynamic integration.
We calculate the free energy as a function of density by integrating the EOS from a reference density to
the density of interest:
\begin{equation} 
F(\rho_1^*)= 
F(\rho_0)
+\int_{\rho_0}^{\rho_1} d\rho
\left\langle \frac{N P(\rho)}{\rho^2}\right\rangle \label{eqn:int_eos}
\end{equation}

\section{Results}

\begin{figure}

\centering
\includegraphics[width=0.45\textwidth]{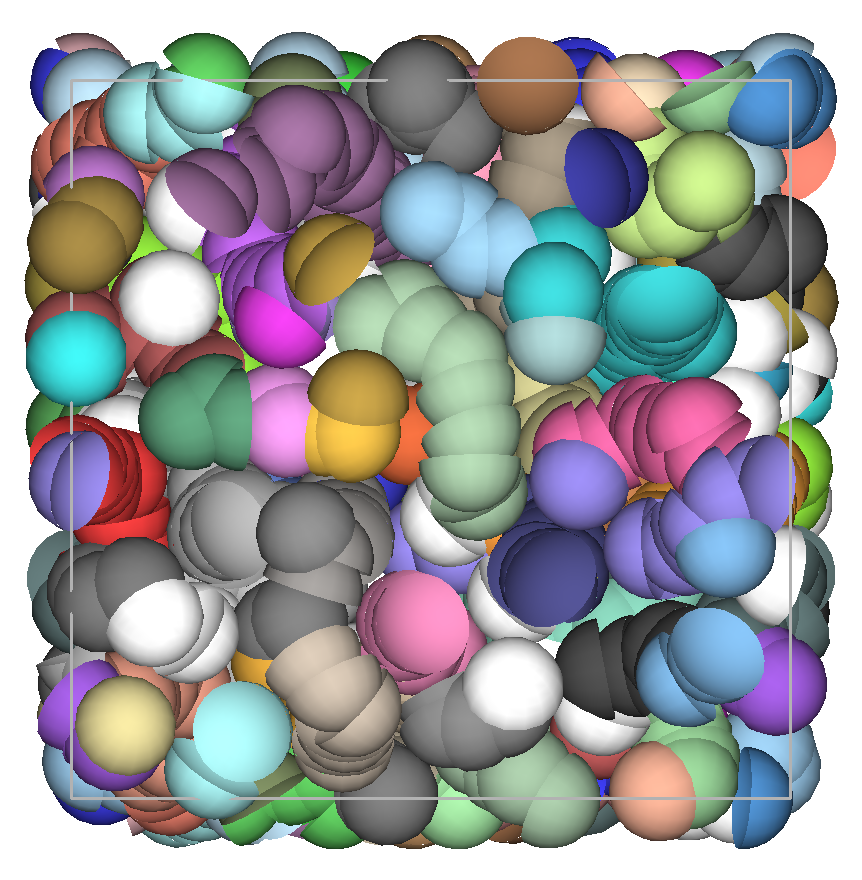}

\caption{ 
The final configuration obtained from simulations at
$P\sigma^3/k_B T=50$ and $D=0.3\sigma$
The colors denote different
stacks.\label{fig:snapshots_worms}}
\end{figure}
\subsection{Stacks}

We perform standard Monte Carlo simulations in the
isobaric-isothermal ensemble (NPT). Fig.~\ref{fig:snapshots_worms}
shows a typical configuration of bowl-shaped particles with $D =
0.3~\sigma$ at $P\sigma^3/k_BT=50$, displaying stacking
behavior typical for the worm-like phase.
\begin{figure}[b!]
\centering
\includegraphics[width=0.49\textwidth]{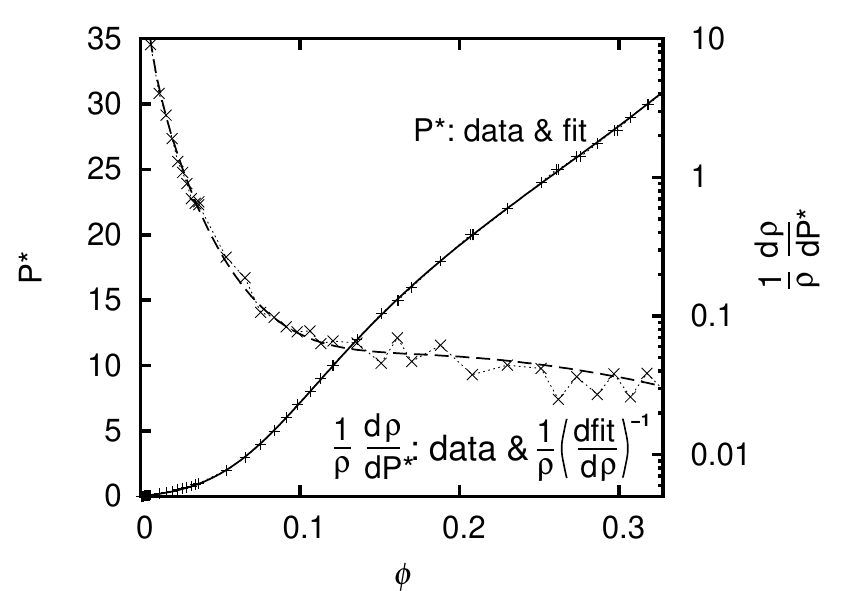}
\caption{The equation of state for bowl-shape particles with $D=0.1\sigma$, reduced pressure $P^*=\beta P\sigma^3$ (left axis), and the reduced compressibility $\frac{1}{\rho}\,\frac{\partial \rho}{\partial P^*}$ on a log scale (right axis) as a function of packing fraction $\phi$. 
The points are data obtained from $NPT$ simulations. The solid line is a fit  to the pressure;
the dashed line is the corresponding reduced compressibility, $\frac{1}{\rho}\big(\frac{\partial \mathsf{fit}(\rho)}{\partial \rho}\big)^{-1}$.
\label{fig:eos}}
\end{figure}
The equation of state (EOS) of the fluid is somewhat peculiar: the pressure as a function of density is not always convex for all densities, although the compressibility does decrease monotonically with packing fraction $\phi$ for $D=0.1\sigma$, see Fig.~\ref{fig:eos}, where
the packing fraction is defined as $\phi=\frac{\pi D}{4}(\sigma^2-D\sigma+\frac{2}{3}D^2)N/V$.
This behavior persist for all $D\leq0.2\sigma$, but for $D\geq 0.25\sigma$ the pressure is always convex. We investigate the origin of these peculiarities using  $g_c(z)$, the positional correlation function along the director of a particle, which includes only the particles in a column around a particle,
as defined in
Eq.~(\ref{eqn:gcz}).
\begin{figure}[b!]
\centering
\includegraphics[width=0.49\textwidth]{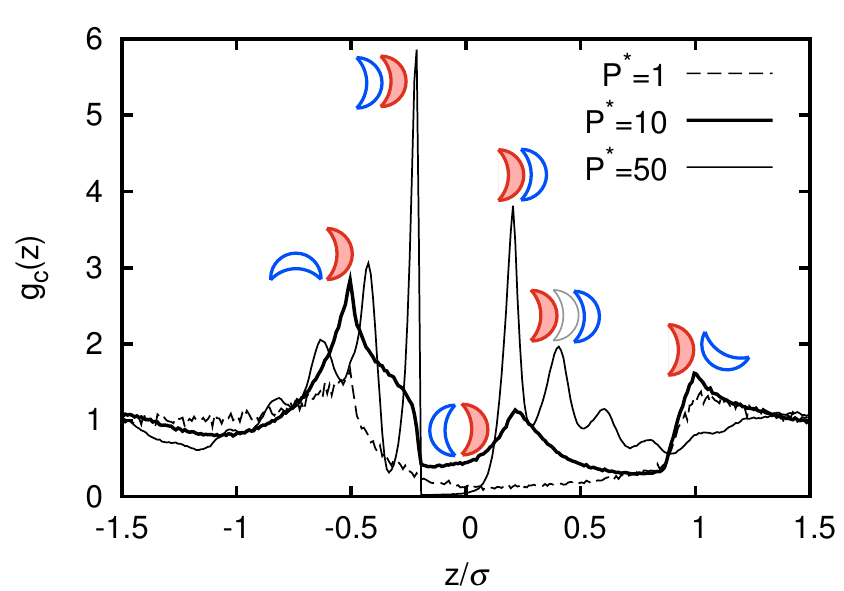}

\caption{The pair correlation function, $g_c(z)$, of a fluid of bowl-shaped particles with $D=0.2\sigma$ as a function of the dimensionless inter-particle distance $z/\sigma$
along the axis of a reference bowl for various reduced pressures $P^*\equiv\beta P\sigma^3$.
Only particles within a cylinder of diameter $\sigma$ around the bowl are considered, as indicated by the subscript `c'.
We show typical two-particle configurations that contribute to $g_c(z)$ for $z/\sigma=-0.5,-0.2,0.2,0.4$ and $1$,
where the filled bowls denote the reference particle, and the open bowls with thick outlines denote the other particle.
\label{fig:gcz}}
\end{figure}
As can be seen from $g_c(z)$ in Fig.~\ref{fig:gcz}, the structure of the fluid changes dramatically as the pressure is increased.
At $P^*\equiv\beta P\sigma^3 =1$, the correlation function is typical for a low density isotropic fluid of hemispherical particles; no effect of the dent  
of the particles is found at low densities.
The only peculiar feature
of $g_c(z)$ for $P^*=1$ is that it is not symmetric around zero, but this is caused by our choice of reference point on the particle (see Fig.~\ref{fig:particles}b), which is located
below the particle if the particle points upwards. In contrast, at $P^*=10$ $g_c(z)$  already shows strong structural correlations.
Most noteworthy is the peak at $z=D$, that shows that
the fluid is forming short stacks of aligned particles. Also, note that the value of $g_c(z)$ is nonzero around $z=0$. This is caused by pairs of bowls that align anti-parallel and form a sphere-like object, as depicted in Fig.~\ref{fig:gcz}.
  Finally, at $P^*=50$ and higher, long worm-like stacks are fully formed and $g_c(z)$ shows multiple peaks at
$z=D n$ for
both positive and negative integer values of $n$. Furthermore, at these pressures, there are no sphere-like pairs, as can be observed from the value of $g_c(0)$.
The formation of stacks explains the peculiar behavior of the pressure: At low densities, the bowls rotate freely, which means that the pressure will be dominated 
by the rotationally averaged excluded volume. The excluded volume of two particles that are not aligned is nonzero, even for $D=0$, and gives rise to 
 the convex pressure which is typical for repulsive particles. As the density increases and the bowls start to form stacks,
the available volume increases, and the pressure increases less than expected, which can even cause the EOS to be concave.
At even higher densities the worm-like stacks are fully formed, and the pressure is again a convex function of density for $D>0$, dominated by the excluded volume of locally aligned bowls. The excluded volume of completely aligned infinitely thin bowls is zero, and, therefore, the pressure increases almost linearly with density for $D=0$ when the stacks are fully formed.

\begin{figure}

\includegraphics[width=0.49\textwidth]{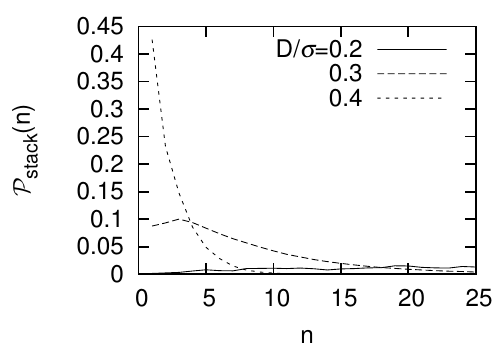}

\caption{The probability, $\mathcal{P}_\text{stack}(n)$, to find a
particle in a stack of size $n$ for $D/\sigma=0.2,0.3$ and $0.4$
and $P\sigma^3/k_BT=50$ 
\label{fig:clust}}
\end{figure}

To
quantify the length of the stacks we calculated the stack
distribution as shown in Fig.~\ref{fig:clust}.
As can be seen from the figure, the length of the stacks is strongly dependent on $D/\sigma$. However, we have found that above a certain threshold pressure
the distribution of stacks is nearly independent of pressure.

\begin{figure}[b]
\centering
\includegraphics[width=0.45\textwidth]{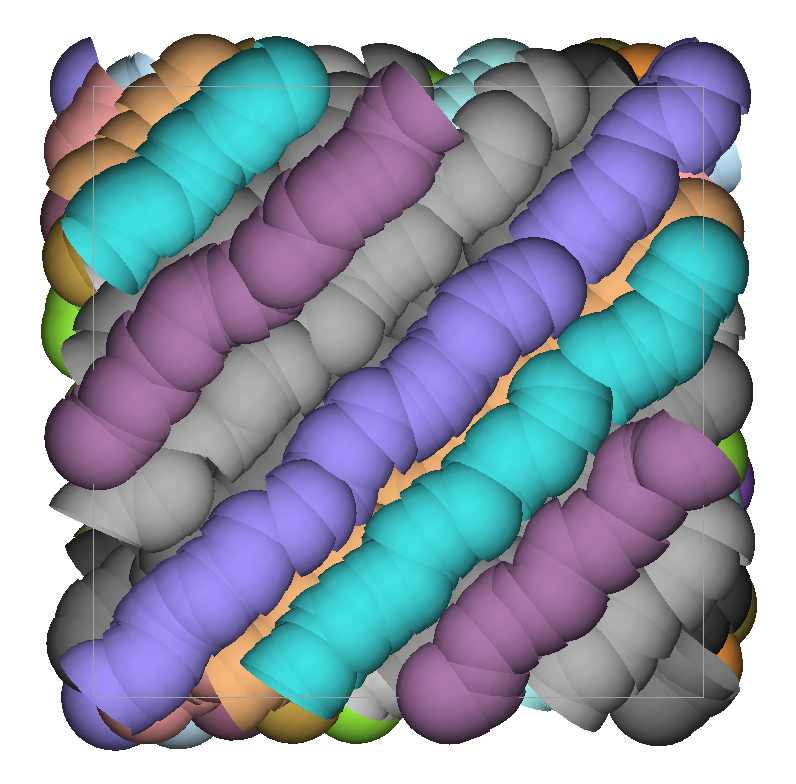}

\caption{
The final configuration of a simulation of bowls with $L=0.1D$ at $P\sigma^3/k_B T=38$.  The gray values denote different
columns.
\label{fig:snapshot_col}}
\end{figure}

We investigated whether the worm-like stacks could spontaneously reorient to form a columnar phase.
We increased the pressure in small steps of 1 $k_BT/\sigma^3$ from well below the fluid--columnar transition to very high pressures, where the system 
was essentially jammed. At each pressure, we
ran the simulation for $4\cdot 10^6$ Monte Carlo cycles, where a cycle consists of $N$ particle and volume moves.
These simulations show that the bowls with a thickness $D\geq 0.25
\sigma$ always remained arrested in the worm-like phase, which is similar to the
experimental observations~\cite{Marechal2010bowls}.
However, for $D/\sigma=0.1$ and $0.2$, we find that
the system eventually transforms into a columnar phase in the
simulations (see Fig.~\ref{fig:snapshot_col}).
 This might be
explained by the fact that the isotropic-to-columnar transition
occurs at lower packing fractions for deeper bowls (smaller $D$),
which facilitates the rearrangements of the particles into stacks
and the alignment of the stacks into the columnar phase.

\subsection{Packing}

\begin{figure}
\centering
\includegraphics[width=0.49\textwidth]{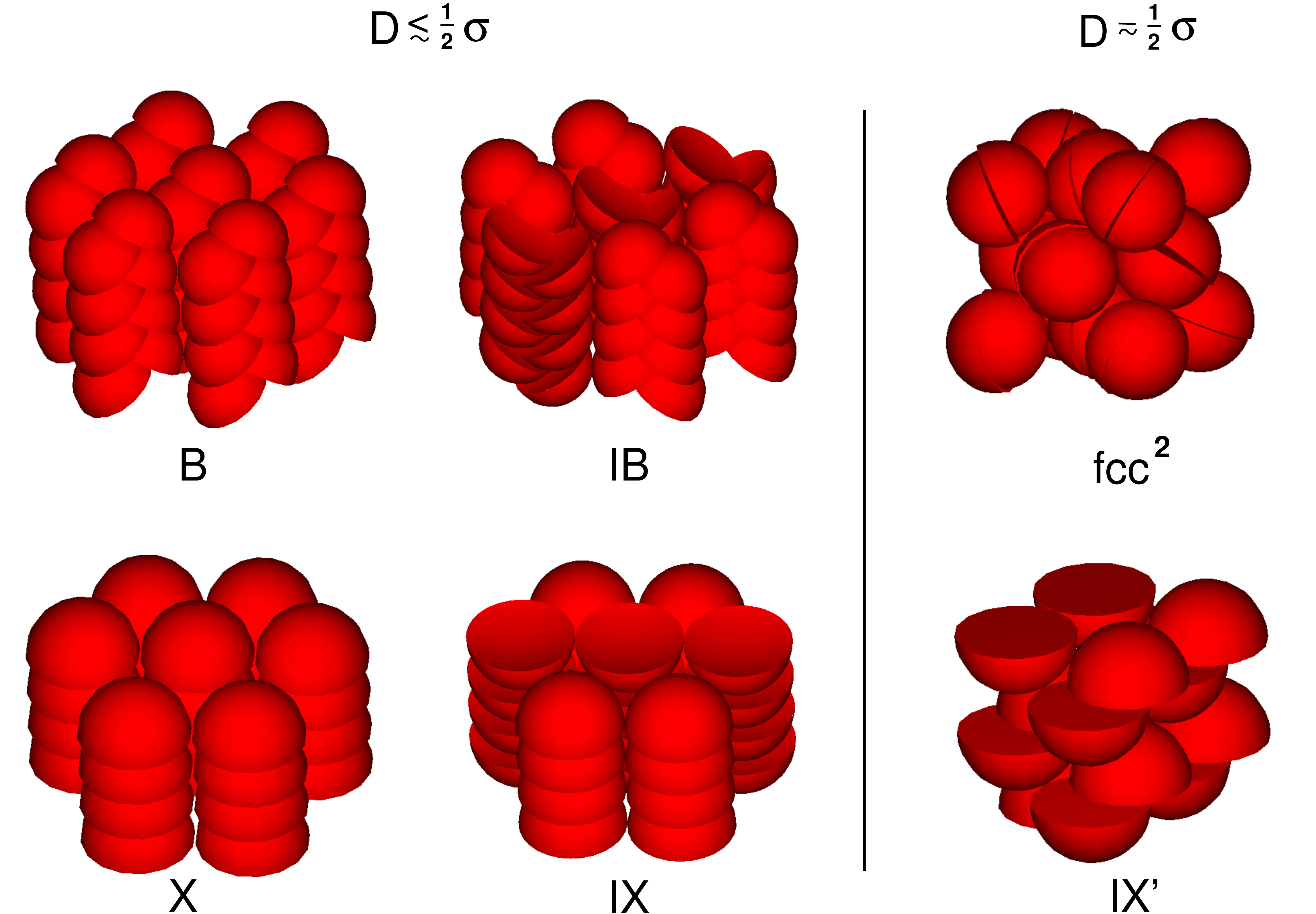}

\caption{The various crystal phases that were considered as possible stable structures. Five of these were found using the pressure annealing method: X, IX, B, IB and IX'. X, IX, B and IB are densely packed structures 
 for $D \lesssim 0.5\,\sigma$ and fcc${}^2$ and IX' are densely packed crystal structures for (nearly) hemispherical bowls
($D \simeq 0.5\,\sigma$).
 \label{fig:crystals}}
\end{figure}

We found six candidate crystal structures, denoted X,IX,IX',B,IB and fcc${}^2$, using the pressure annealing method.
Snapshots of a few unit cells of these crystal phases are shown in Fig.~\ref{fig:crystals}.
We will describe these crystal structures using the order parameters $S_1$, that measures alignment of the particles, and the nematic order parameter ($S_2$), that 
is nonzero for both parallel and anti-parallel configurations.
   Crystal structure X has
$S_1\simeq 1$ and $S_2\simeq 1$, and the particles are stacked head to toe in columns.
The lattice vectors are
\begin{equation}
\begin{array}{c}
\displaystyle
\mathbf{a_1}=\sigma \hat{x} \qquad \mathbf{a_2}=D \hat{z} \\[1.0em]
\displaystyle
 \mathbf{a_3}=\frac{\sigma}{2}\hat{x} + \frac{1}{2} \sqrt{\sigma^2-D^2+2\sigma \sqrt{\sigma^2-D^2} }\;\hat{y}+\frac{D}{2} \hat{z},
\end{array}
\end{equation}
and the density is
\begin{equation}
\rho\sigma^3=\left[\frac{D \sigma}{2} \sqrt{\sigma^2-D^2+2\sigma \sqrt{\sigma^2-D^2} }\right]^{-1}.
\end{equation}

The order parameters of the second crystal structure, are $S_1\simeq 0$ and $S_2\simeq 1$, which is caused by the fact that
half of the particles point upwards, and the other half downwards.
 Further investigation shows that there are two phases with $S_2 \simeq 1$ and $S_1 \simeq0$: one at low $D$ (IX) and one at $D\simeq \sigma/2$ (IX'). 
The structure within the columns of the first (IX) of these two structures is the same as for the X structure,
but one half of these columns are upside down, like in the inverted
columnar phase (in fact, the IX crystal melts into the inverted columnar phase). 
The lattice vectors of crystal structure IX are
\begin{equation}
\begin{array}{c}
\displaystyle
\mathbf{a_1}=\sigma \hat{x} \qquad \mathbf{a_2}=D \hat{z} \\[1.0em]
\displaystyle
 \mathbf{a_3}=\frac{\sigma}{2}\hat{x} + \frac{1}{2} \sqrt{3\sigma^2-4 D^2}\;\hat{y},
\end{array}\label{eqn:IXa}
\end{equation}
and the density is
\begin{equation}
\rho\sigma^3=\left[\frac{D \sigma}{2} \sqrt{3\sigma^2-4 D^2}\right]^{-1}.\label{eqn:IXd}
\end{equation}
The columns in the IX crystal are arranged in such a way that the rims
of the bowls can interdigitate.
The IX' crystal can be obtained from the IX phase at $D=\sigma/2$ by shifting every other layer by some distance perpendicular to the columns, such that the particles in these layers fit into
the gaps in the layers below or above. In this way a higher density than Eq.~(\ref{eqn:IXd}) is achieved. 
The columns of the third crystal phase (B) resemble braids with alternating tilt direction
 of the particles within each column. 
Because of this tilt $S_1$ and $S_2$ have values between 0 and 1, that depend on $D$.
Furthermore, the inverted braids structure (IB), that has $0<S_2<1$ and $S_1=0$, can be obtained
by flipping one half of the columns of the braid-like phase (B) upside down.
These braid-like columns piece together in such a way that the particles are interdigitated. In other words, this phase is related to the B phase in exactly the same way as the IX phase is
related to the X phase.
\begin{figure}[b!]
\centering
\includegraphics[width=0.49\textwidth]{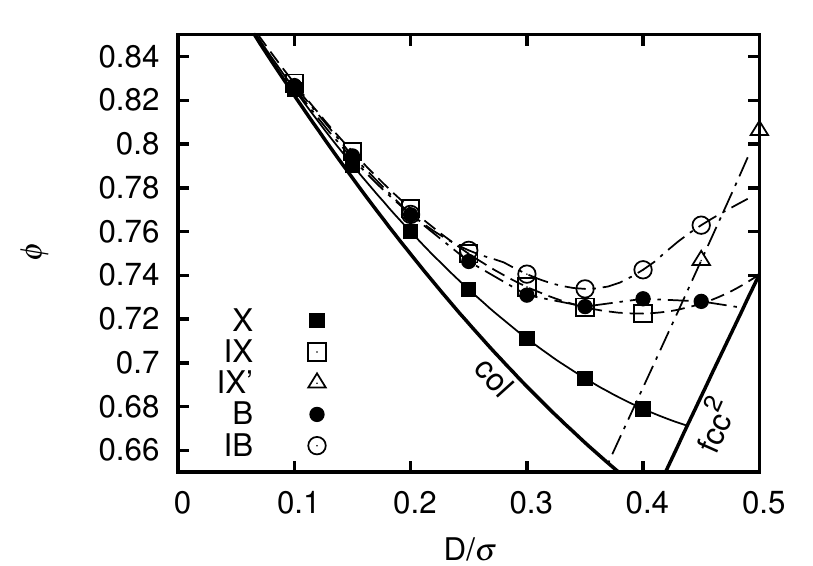}

\caption{Packing diagram: maximum packing fraction ($\phi$) of various crystal
phases as a function of the thickness
($D$) of the bowls. The points are the results of the pressure annealing simulations. The thin dot-dashed lines are obtained from the pressure annealing results by slowly increasing or decreasing $D$ as described in Sec.~\ref{sec:cryst}, except for the IX phase (thin dashed line with open squares) and the X phase (thin solid line with filled squares),
for which the packing fraction can be expressed analytically.
 The thick lines denote the packing fractions of the perfect hexagonal columnar phase (col) and the
paired fcc phase (fcc${}^2$). Any points that lie below these lines are expected to be thermodynamically unstable (see text).
\label{fig:packing}}
\end{figure}
Finally, in the paired face-centered-cubic (fcc${}^2$) phase, pairs of
hemispheres form sphere-like objects that can rotate freely
and that are located at the lattice positions of an fcc crystal.
The density at close packing is $2\sqrt{2}/\sigma^3$, \emph{i.e.}  twice the density of fcc.

In Fig.~\ref{fig:packing} the results of the pressure annealing method are shown, along with the known packing fraction of the perfect hexagonal columnar phase (col). Since the columnar phase has positional degrees of freedom and the fcc${}^2$ phase has rotational degrees of freedom,
we expect these phases to have a higher entropy (lower free energy) than any crystal
phase with the same or lower maximum packing fraction whose degrees of freedom have all been frozen out. Therefore,
 any crystal structure with a packing fraction below the thick lines in Fig.~\ref{fig:packing} is most likely thermodynamically unstable.
At first, we were unable to find the fcc${}^2$ 
 using the pressure annealing method as described in 
Sec.~\ref{sec:cryst}. However, if we increase the pressure slowly to 100$k_BT/\sigma^3$ in simulations of 12 particles, we did
observe the fcc${}^2$ phase for hemispherical particles ($D=\sigma/2$). In these simulations at finite pressure, it is important to constrain the length of all box vectors such that they remain larger
than say $1.5\sigma$.
Otherwise the box will become extremely elongated, such that the particles can interact primarily with their own images. 
When a particle interacts with it is neighbors, the Gibbs free energy $G=F+PV$ decreases, because the volume decreases without any decrease in entropy due to restricted translational motion 
(if a particle moves, its image moves as well, so a particle translation will never cause overlap of the particle with its image).
The decrease in Gibbs free energy is of course an extreme finite size effect, which should be avoided if we wish to predict the equilibrium phase behavior.
For the pressure annealing simulations at very high pressures, these effects are not important, because the entropy term in the Gibbs free energy is small compared to $PV$.
\begin{table}[b!]
\centering
{\setlength\tabcolsep{1em}\begin{tabular}{l|l|l|l|l}
phase & $D/\sigma$ & $\rho_\text{fluid}\sigma^3$&$\rho_\text{col}\sigma^3$ & $f_\text{diff}$ \\
\hline
fluid--col & 0 & 1.461&4.679 & 7.33272 \\
\multicolumn{3}{c}{} \\
phases & $D/\sigma$ & $\phi_\text{fluid}$& $\phi_\text{col}$ & $f_\text{diff}$ \\
\hline
fluid--col & 0.1 & 0.1780& 0.2848 & 3.2630(7) \\
fluid--col & 0.2 & 0.3116& 0.4674 & 3.268(2) \\
fluid--col & 0.3 & 0.3760& 0.5193 & 3.802(1) \\
fluid--inv col & 0.3 & 0.3760& 0.5193 & 3.8155(8) \\
fluid--col & 0.4 & 0.4440& 0.5772 & 5.843 \\
\end{tabular}}
\caption{Free energy differences, $f_\mathrm{diff}\equiv (F_\text{col}(\rho_\text{col})-F_\mathrm{fluid}(\rho_\text{fluid})/(N k_B T)$, between the (inverted) columnar phase at density $\rho_\text{col}$ or packing fraction $\phi_\text{col}$ and the fluid phase at $\rho_\text{fluid}$ or $\phi_\text{fluid}$. 
In the column ``phases'', ``col'' denotes the columnar phase and inverted columnar phase is abbreviated to ``inv col''.
\label{tbl:col_fs}}
\end{table}
We did not attempt to find the columnar phase using the modified pressure annealing method, as we were only interested in finding candidate crystal structures.
Furthermore, the columnar phase was already found in more standard simulations with a larger number of particles.

\subsection{Free energies}

In order to determine the regions of the stability of the fluid, the columnar phase and the six crystal phases, 
we calculated the free energies of all phases as
explained in the Methods section. The results of the reference free energy calculations are shown in Tbls.~\ref{tbl:col_fs} and \ref{tbl:fs}.

\begin{table}[b]
\centering
{\setlength\tabcolsep{1.5em}\begin{tabular}{l|l|l|l}
phase & $D/\sigma$ & $\phi$ & $f_\mathrm{exc}$ \\
\hline
IX & 0.3 & 0.6669 & 15.505(4) \\
IB & 0.3 & 0.6971 & 18.407(3) \\
IX & 0.4 & 0.6177 & 12.52(1) \\
IB & 0.4 & 0.6170 & 13.195(2) \\
IX & 0.45 & 0.6768 & 17.918(2) \\
IB & 0.45 & 0.6662 & 14.9873(4) \\
fcc${}^2$ & 0.45 & 0.6192 & 12.8591(5) \\
IX' & 0.45 & 0.6950 & 18.170(5) \\
fcc${}^2$ & 0.5 & 0.5455 & 8.7673(7) \\
IX' & 0.5 & 0.5597 & 10.854(3) \\
\end{tabular}}
\caption{Excess free energies, $f_\mathrm{exc}\equiv (F-F_\mathrm{id})/(N k_B T)$, of the various crystal phases, where $F_\mathrm{id}$ is the ideal gas free energy.
The various crystal phases are labeled as in Fig.~\ref{fig:crystals}.
\label{tbl:fs}}
\end{table}
We find that
the columnar phase with all the particles pointing in the same
direction is more stable than the inverted columnar phase, where half of the
columns are upside down. However, the free energy difference
between the two phases is only $0.013\pm 0.002 k_B T$ per particle at
$\phi=0.5193$ and $D=0.3\sigma$. Based on this small free energy difference
we do not expect polar ordering to occur spontaneously.
Similar conclusions, based on direct simulations, were already drawn in
Ref.~\cite{simulation-bowls}.

\begin{figure}[t!]
\centering
\includegraphics[width=0.49\textwidth]{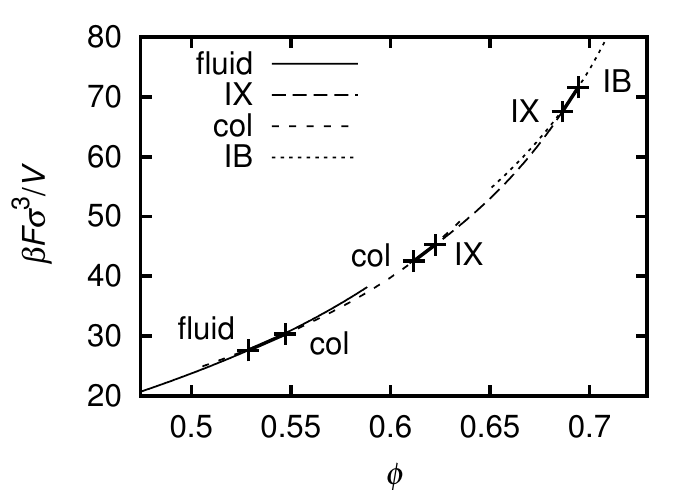}
\caption{Dimensionless free energies $\beta F\, \sigma^3/V$ for hard bowls with $L=0.3\sigma$ and the fluid-columnar, columnar-IX and IX-IB coexistences, which were calculated using  common tangent constructions. The columnar phases is denoted ``col''.
The irrelevant free energy offset is defined in such a way that the free energy of the ideal gas reads
$\beta F/V=\rho (\log(\rho\sigma^3)-1)$.
The free energies of the various phases are so close, that they are almost indistinguishable.
\label{fig:f_of_rho}}
\end{figure}
The densely-packed crystal structures in Fig.~\ref{fig:crystals} at $D \lesssim 0.3$,
the worm-like fluid phase (Fig.~\ref{fig:snapshots_worms}) and the columnar phase (Fig.~\ref{fig:snapshot_col}) show striking similarity
in the
local structure: in all these phases the bowls are stacked on top of each other, such that (part of) one bowl fits into the dent
of another bowl. As a result, the free energies and pressures of the various phases,
are often almost indistinguishable near coexistence. For this reason it was sometimes difficult to determine the coexistence densities for $D<0.3\sigma$. Exemplary 
free energy curves for the various stable phases consisting of bowls with $D=0.3\sigma$ are shown in Fig.~\ref{fig:f_of_rho}.

\subsection{Phase diagram}

\begin{figure}[b]
\centering
\includegraphics[width=0.49\textwidth]{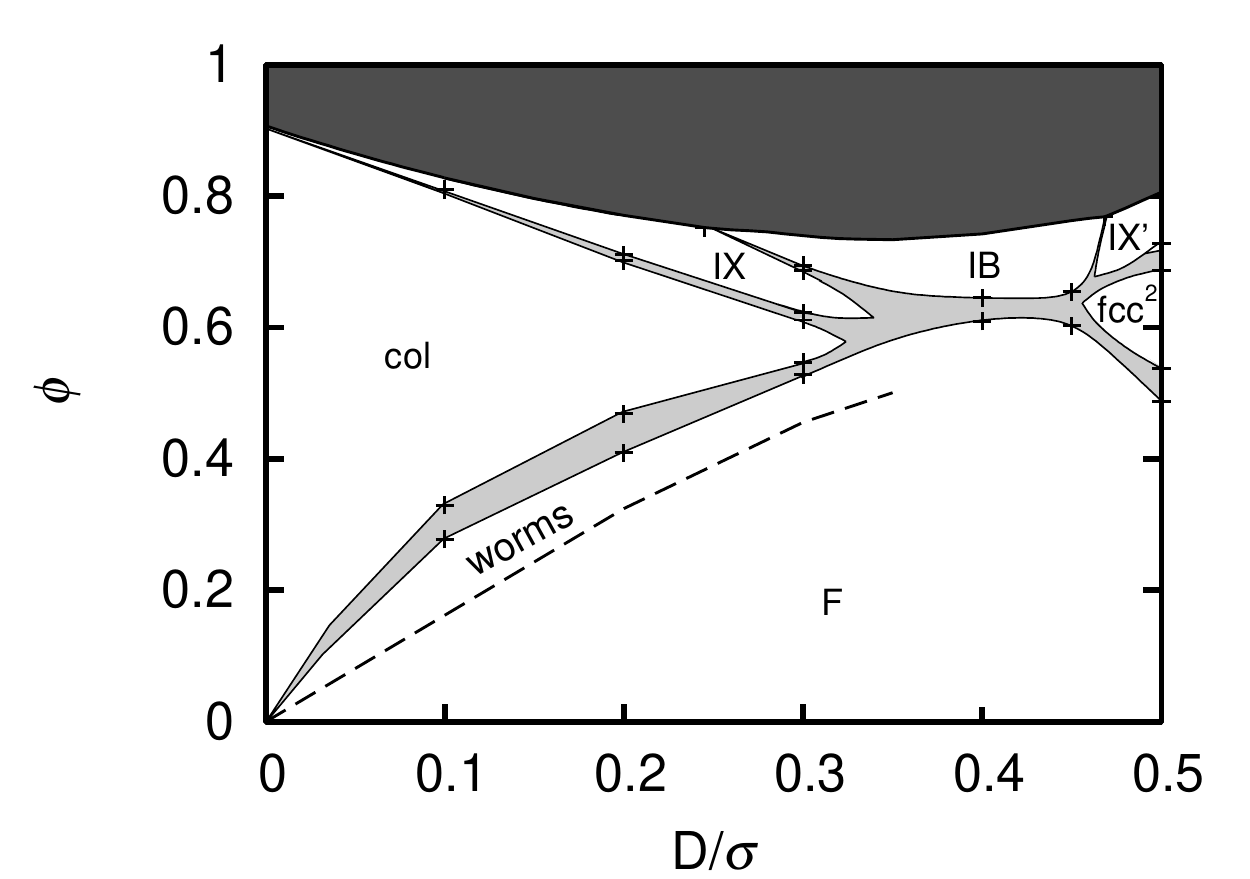}

\caption{Phase diagram in the packing fraction ($\phi$) versus thickness
($D$) representation. The light gray areas are coexistence areas, while the state points in the dark gray area are inaccessible since they lie
above the close packing line. IX, IB, IX' and fcc${}^2$ denote the crystals as shown in Fig.~\ref{fig:crystals}, ``F'' is the fluid and ``col'' is the columnar phase.
 The lines are a guide to the eye. Worm-like stacks were found in the area marked ``worms'' bounded from below by the dashed line. This line denotes the probability to find a particle in a cluster that consists of more than two particles, $\mathcal{P}_\text{stack}(n>2)=1/2$.
\label{fig:ph-dia_shells}}
\end{figure}

In Fig.~\ref{fig:ph-dia_shells}, we show the phase diagram in
the packing fraction $\phi$ - thickness $D/\sigma$ representation.
The packing fraction is defined as $\phi=\frac{\pi
D}{4}(\sigma^2-D\sigma+\frac{2}{3}D^2)N/V$.
For $D/\sigma\leq0.3$, we find an isotropic-to-columnar phase
transition at intermediate densities, which resembles the phase
diagram of thin hard discs~\cite{Veerman_Frenkel}. However, the
fluid-columnar-crystal triple point for discs is at  a
thickness-to-diameter ratio of about $L/\sigma \sim 0.2-0.3$,
while in our case the triple point is at about $D/\sigma \sim
0.3-0.4$. The shape of the bowls stabilizes the
columnar phase compared to the fluid and the crystal phase. 
We find four stable crystal phases IX, IB, IX' and fcc${}^2$, while we had six candidate crystals.
The two phases that were not stable are the X and B crystals, which are very similar to the stable IX and IB crystals respectively, except that X and B have considerable lower close packing densities.
Therefore, one could have expected these phases to be unstable. 
On the other hand, we observe from the phase diagram, that IX is stable at intermediate densities for $0.25\sigma < D< 0.45\sigma$,
while IB packs better than IX.
In other words, stability can not be inferred from small differences in packing densities.

\begin{table}[t!]
\centering
\noindent\begin{tabular*}{0.49\textwidth}{@{\extracolsep{\fill}}*{7}{c@{\extracolsep{\fill}}}}
$D/\sigma$ & phase 1 & phase 2 & $\rho_1\sigma^3$ & $\rho_2\sigma^3$ & $\beta P \sigma^3$ & $\mu^*$ \\
\hline
0 & fluid & col & 4.083 & 4.824  & 26.11 & 15.22 \\
\hline 
\end{tabular*}\\[1em]
\begin{tabular*}{0.49\textwidth}{@{\extracolsep{\fill}}*{7}{c@{\extracolsep{\fill}}}}
$D/\sigma$ & phase 1 & phase 2 & $\phi_1$ & $\phi_2$ & $\beta P d^3$ & $\mu^*$ \\
\hline
0.1 & fluid & col & 0.2778 & 0.3297 & 26.35 & 15.59\\
0.1 & col & IX & 0.8095 & 0.8104 & $2.7\!\cdot\!10^3$ & -\\
0.2 & fluid & col & 0.4096 & 0.4688 & 27.23 & 16.68\\
0.2 & col & IX & 0.7021 & 0.7108 & 325 & -\\
0.3 & fluid & col & 0.5286 & 0.5472 & 49.52 & 26.13\\
0.3 & col & IX & 0.6864 & 0.6944 & 281.4 & 91.03\\
0.3 & IX & IB & 0.6117 & 0.6226 & 110.9 & 44.92\\
0.4 & fluid & IB & 0.6098 & 0.6455 & 105.9 & 51.06\\
0.45 & fluid & IB & 0.6026 & 0.6545 & 87.92 & 46.90\\
0.5 & fluid & fcc${}^2$ &  0.4878 & 0.5383 & 28.34 & 22.10\\
0.5 & fcc${}^2$ & IX' &  0.6870 & 0.7278 & 139.2 & 67.36\\
\hline
\end{tabular*}\hspace{0.07\textwidth}%
\caption{Reduced densities, pressures and chemical potentials $\mu^*=\beta\mu-\ln(\Lambda_t^3 \Lambda_r / \sigma^3)$ of 
the coexisting phases for hard bowl-shaped particles with thickness $D$.
\label{tbl:coexes}}
\end{table}

Almost all coexistence densities were calculated
by employing the common tangent construction
to the free energy curves,
except for the col--IX coexistence at $D=0.1\sigma$ and $0.2\sigma$. At these values of $D$ the transition occurs at very high pressures, while the
free energy of the columnar phase is calculated at the fluid--col transition, which occurs at a low pressure. To get a value for the free energy
of the columnar phase we would have to integrate the equation of state up to these high pressures, accumulating integration errors.
Furthermore, we expect the coexistence to be rather thin, which would further complicate the calculation.
So, instead we just ran long variable box shape $NPT$ simulations to see at which pressure the IX phase melts into the inverted columnar phase.
As the free energy difference between the inverted columnar phase and the columnar phase is small,
we assume that this is the coexistence pressure for the col--IX transition, although technically it is only a lower bound.
The density of the columnar phase at this pressure is determined using a local fit of the equation of state.
All coexistences are tabulated in Tbl.~\ref{tbl:coexes}.
We draw a tentative line in the phase diagram to mark
the transition from a structureless fluid to a worm-like fluid \emph{i.e.} a fluid with many stacks.
In a dense but structureless fluid, stacks of size 2 are quite probable, but larger stacks occur far less frequently. We calculate the probability to find a particle in a stack that contains more than 2 particles $\mathcal{P}_\text{stack}(n>2)=1-\mathcal{P}_\text{stack}(1)-\mathcal{P}_\text{stack}(2)$ and define the worm-like phase by the criterion $\mathcal{P}_\text{stack}(n>2)\geq 1/2$ in Fig.~\ref{fig:ph-dia_shells}. We do not imply that the transition to the worm-like phase is a true thermodynamic phase transition; the transition is rather 
continuous. The type of stacks in the fluid changes from worm-like for $D=0.3\sigma$ to something resembling 
the columns in the braid-like crystals B and IB (see Fig.~\ref{fig:crystals}) for $D=0.4\sigma$.
Therefore, the region of stability worm-like phase was ended at $D=0.35\sigma$, where there are similar amounts of braid-like and worm-like stacks.

\section{Summary and discussion}

We have studied the phase behavior of hard
bowls in Monte Carlo simulations.
We find that the bowls have a strong tendency to
form stacks, but the stacks are bent and not aligned.
We measured the equation of state and the compressibility in Monte Carlo $NPT$ simulations.
The pressure we obtained from these simulations is concave for some range of densities for deep bowls.
This is due to the increase in free volume when large stacks form.
Using 
$g_c(z)$, the pair correlation function along the direction vector, 
we showed that the concavity of the pressure coincides with a dramatic change in structure from a homogeneous fluid to the worm-like fluid.
We measured the three-dimensional stack length distribution in the simulations.
When the pressure is increased slowly, the deep
bowls spontaneously
order into a columnar phase in our simulations. This poses severe restrictions on
the thickness of future bowl-like mesogens (molecular or colloidal), which are designed to
easily order into a globally aligned lyotropic columnar phase. We
determined the phase diagram using free energy calculations 
for a particle shape ranging from an
infinitely thin bowl to a solid hemisphere. We find that the
columnar phase is stable for $D\leq0.3\sigma$ at intermediate
packing fractions. In addition, we show using free energy
calculations that the stable columnar phase possesses polar order.
However, the free energy penalty for flipping columns upside down
is very small, which makes it hard to achieve complete polar
ordering in a spontaneously formed columnar phase of bowls. 

\section*{Acknowledgments}

We thank Rob Kortschot, Ahmet Demir\"ors and Arnout Imhof for useful discussions.
Financial support is acknowledged from an NWO-VICI grant and from the High Potential Programme of Utrecht University.

\appendix

\section{Overlap algorithm}

The overlap algorithm for our bowls checks whether the surfaces of two bowls intersect.
Fig.~\ref{fig:particles} shows that the surface of the bowl consists of two parts. Part $p$ of the surface contains the part of the surface of the sphere of radius $R_p$, within an angle $\theta_p$ from the $z$-axis, where $p=1$ denotes the smaller sphere and the larger sphere is labeled $p=2$.
We set $\theta_1=\pi/2$, to get a hemispherical outer surface.
The edges of both surfaces have to coincide, such that our particles have a closed surface.
Using this restriction
$L$, $\theta_2$ and $R_2$ can all be expressed in terms of the radius of the smaller sphere, $R_1$, and 
 the thickness of the bowl $D$, in the following way:
{\setlength\arraycolsep{1pt}
\begin{eqnarray}
R_2 & = & R_1+\frac{D^2}{2(R_1-D)}\\
\theta_2 & = & \arcsin(R_1/R_2)\\
L & = & R_2 \cos(\theta_2).
\end{eqnarray}
}

Overlap occurs if either of the two parts of the surface of a bowl overlaps with either of the two parts of another bowl.
So we have to check four pairs of
infinitely thin (and not necessarily hemispherical) bowls, labeled $i$ and $j$, for overlap.
The algorithm for two such surfaces that are equal in shape was already implemented by He and Siders~\cite{He1990UFOs}
as part of their overlap algorithm for their ``UFO'' particles, which are defined as the intersection between two spheres.
An equivalent overlap algorithm was used by Cinacchi and Duijneveldt~\cite{Cinacchi2010} to simulate infinitely thin 
 contact lense-like particles, but the overlap algorithm was not described explicitly.
We can not use one of these algorithms, since the two parts of the surface of our particle are unequal in shape. 
Therefore, we implemented a slightly different version of the overlap algorithm, which we describe in the remainder of this section.
In our overlap algorithm, the existence of a overlap or intersection between two infinitely thin bowls is checked in three steps.
\begin{itemize}
\item First, we check whether the full surfaces of the spheres intersect, i.e. $|R_i-R_j|<r_{ij}\equiv|\mathbf{r}_j-\mathbf{r}_i|<R_i+R_j$.
 If this intersection does not exist, there is no overlap, otherwise we proceed to the next step.
\item Secondly, we determine the intersection of the surface of each sphere with the other bowl.
The intersection of bowl $i$ with the sphere of bowl $j$ exists if
\begin{equation}
|\omega_{ij}+\zeta \phi_{ij}| < \theta_i \label{eqn:sphere_bowl}\\
\end{equation}
for $\zeta=1$ or $-1$,
where
\begin{eqnarray}
\cos(\phi_{ij})& =& \frac{R_i^2-R^2_j+r_{ij}^2}{2 r_{ij} R_i}\ \mathrm{and}\\
\cos(\omega_{ij})&=& \frac{\mathbf{u}_i \cdot \mathbf{r}_{ij}}{r_{ij}}.
\end{eqnarray}
see Fig.~\ref{fig:calc_shells}a.
\begin{figure}[b!]
\centering
\includegraphics[width=0.49\textwidth]{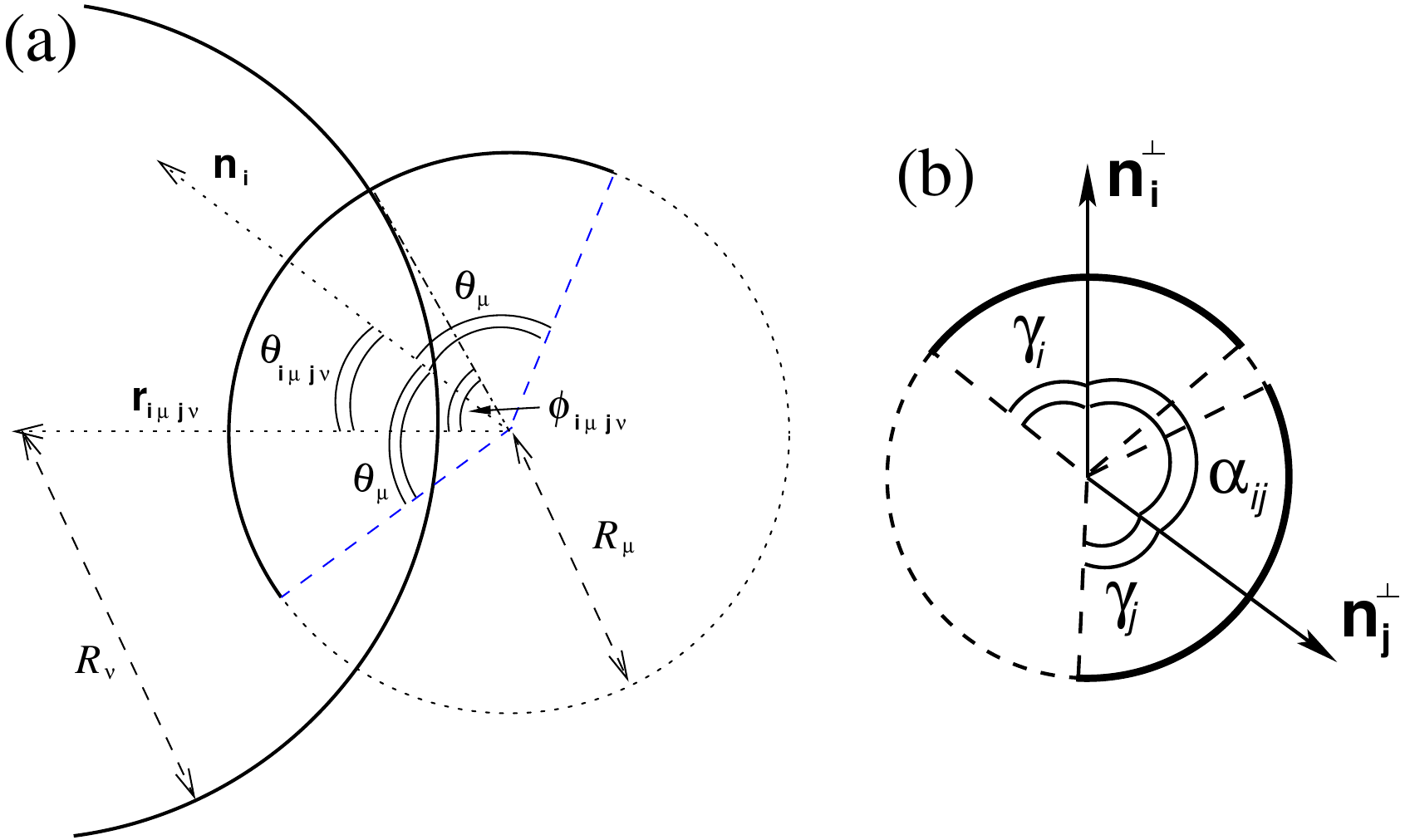}
\caption{
The relevant lengths and angles which are used in the first and second steps (a) and in the third step (b) of the overlap algorithm.
Shown are bowl $i$ and (part of) the sphere of bowl $j$ (a),
the arcs of $i$ and $j$ and the circular intersection of the spheres of $i$ and $j$ (b).
In (a) $\mathbf{r}_{ij}$ lies in the plane, while the plane of view in (b) is perpendicular to $\mathbf{r}_{ij}$.
In this case, the sphere of particle $j$ overlaps with bowl $i$, but the arcs do not overlap, so
particle $i$ and particle $j$ do not overlap.
\label{fig:calc_shells}}
\end{figure}
This intersection is an arc, which is part of the circle that is the intersection between the two spheres.
If in fact this arc is a full circle and the other particle has a nonzero intersection, the particles overlap.
This is the case when Eq.~(\ref{eqn:sphere_bowl}) holds for $\zeta=1$ \emph{and} $\zeta=-1$.
 If, on the contrary, either of the two arcs does not exist, there is no overlap. Otherwise, if both arcs exist, but neither of them is a full circle, proceed to the next step.
\item Finally, if the two arcs overlap there is overlap, otherwise the particles do not overlap. The arcs overlap if
\begin{equation}
|\alpha_{ij}|<|\gamma_i|+|\gamma_j|,\label{eqn:arc_overlap}
\end{equation}
where
\begin{eqnarray}
\cos(\alpha_{ij})&=&\frac{ \mathbf{n}_i^\perp \cdot \mathbf{n}_j^\perp}{|\mathbf{n}_i^\perp| |\mathbf{n}_j^\perp|} \\
\cos(\gamma_i)&=& \frac{\cos(\theta_i)-\cos(\phi_{ij})\cos(\omega_{ij})}{\sin(\phi_{ij})\sin(\omega_{ij})},
\end{eqnarray}
where $\mathbf{n}_i^\perp=\mathbf{n}_i-(\mathbf{r}_{ij}\cdot\mathbf{n}_i)\mathbf{r}_{ij}/r^2_{ij}$ and
the expressions for $\gamma_j$ and $\mathbf{n}_j^\perp$ are equal to the expressions for $\gamma_i$ and $\mathbf{n}_i^\perp$ with $i$ and $j$ interchanged. The arcs together
with the relevant angles are drawn in Fig.~\ref{fig:calc_shells}b.
\end{itemize}
The inequalities (\ref{eqn:sphere_bowl}) and (\ref{eqn:arc_overlap}) are expressed in cosines and sines using some simple trigonometry.
In this way no inverse cosines need to be calculated during the overlap algorithm.

For $D=0.5\sigma$ the bottom surface is a disk rather than an infinitely thin bowl. So the overlap check consists of bowl--bowl, bowl--disc and disc--disc overlap checks.
For brevity, we will not write down the bowl--disk overlap algorithm, but it can be implemented in a similar way as the algorithm 
for bowl--bowl overlap described above. The disk--disk overlap algorithm was already implemented by Eppenga and Frenkel~\cite{Eppenga}.

\bibliography{bowls}

\begin{thebibliography}{32}%
\makeatletter
\providecommand \@ifxundefined [1]{%
 \@ifx{#1\undefined}
}%
\providecommand \@ifnum [1]{%
 \ifnum #1\expandafter \@firstoftwo
 \else \expandafter \@secondoftwo
 \fi
}%
\providecommand \@ifx [1]{%
 \ifx #1\expandafter \@firstoftwo
 \else \expandafter \@secondoftwo
 \fi
}%
\providecommand \natexlab [1]{#1}%
\providecommand \enquote  [1]{``#1''}%
\providecommand \bibnamefont  [1]{#1}%
\providecommand \bibfnamefont [1]{#1}%
\providecommand \citenamefont [1]{#1}%
\providecommand \href@noop [0]{\@secondoftwo}%
\providecommand \href [0]{\begingroup \@sanitize@url \@href}%
\providecommand \@href[1]{\@@startlink{#1}\@@href}%
\providecommand \@@href[1]{\endgroup#1\@@endlink}%
\providecommand \@sanitize@url [0]{\catcode `\\12\catcode `\$12\catcode
  `\&12\catcode `\#12\catcode `\^12\catcode `\_12\catcode `\%12\relax}%
\providecommand \@@startlink[1]{}%
\providecommand \@@endlink[0]{}%
\providecommand \url  [0]{\begingroup\@sanitize@url \@url }%
\providecommand \@url [1]{\endgroup\@href {#1}{\urlprefix }}%
\providecommand \urlprefix  [0]{URL }%
\providecommand \Eprint [0]{\href }%
\@ifxundefined \urlstyle {%
  \providecommand \doi  [0]{\begingroup \@sanitize@url \@doi}%
  \providecommand \@doi [1]{\endgroup \@@startlink {\doibase
  #1}doi:\discretionary {}{}{}#1\@@endlink }%
}{%
  \providecommand \doi  [0]{doi:\discretionary{}{}{}\begingroup
  \urlstyle{rm}\Url }%
}%
\providecommand \doibase [0]{http://dx.doi.org/}%
\providecommand \Doi [0]{\begingroup \@sanitize@url \@Doi }%
\providecommand \@Doi  [1]{\endgroup\@@startlink{\doibase#1}\@@Doi}%
\providecommand \@@Doi [1]{#1\@@endlink}%
\providecommand \selectlanguage [0]{\@gobble}%
\providecommand \bibinfo  [0]{\@secondoftwo}%
\providecommand \bibfield  [0]{\@secondoftwo}%
\providecommand \translation [1]{[#1]}%
\providecommand \BibitemOpen [0]{}%
\providecommand \bibitemStop [0]{}%
\providecommand \bibitemNoStop [0]{.\EOS\space}%
\providecommand \EOS [0]{\spacefactor3000\relax}%
\providecommand \BibitemShut  [1]{\csname bibitem#1\endcsname}%
\bibitem [{\citenamefont {Rabe}\ \emph {et~al.}(2007)\citenamefont {Rabe},
  \citenamefont {Dawber}, \citenamefont {Lichtensteiger}, \citenamefont {Ahn},\
  and\ \citenamefont {Triscone}}]{FerroApp}%
  \BibitemOpen
  \bibfield  {author} {\bibinfo {author} {\bibfnamefont {K.~M.}\ \bibnamefont
  {Rabe}}, \bibinfo {author} {\bibfnamefont {M.}~\bibnamefont {Dawber}},
  \bibinfo {author} {\bibfnamefont {C.}~\bibnamefont {Lichtensteiger}},
  \bibinfo {author} {\bibfnamefont {C.~H.}\ \bibnamefont {Ahn}}, \ and\
  \bibinfo {author} {\bibfnamefont {J.-M.}\ \bibnamefont {Triscone}},\ }in\
  \href@noop {} {\emph {\bibinfo {booktitle} {Physics of Ferroelectrics}}}\
  (\bibinfo  {publisher} {Springer Berlin / Heidelberg},\ \bibinfo {year}
  {2007})\ pp.\ \bibinfo {pages} {1--30}\BibitemShut {NoStop}%
\bibitem [{\citenamefont {Ricci}\ \emph {et~al.}(2008)\citenamefont {Ricci},
  \citenamefont {Berardi},\ and\ \citenamefont {Zannoni}}]{simulation-bowls}%
  \BibitemOpen
  \bibfield  {author} {\bibinfo {author} {\bibfnamefont {M.}~\bibnamefont
  {Ricci}}, \bibinfo {author} {\bibfnamefont {R.}~\bibnamefont {Berardi}}, \
  and\ \bibinfo {author} {\bibfnamefont {C.}~\bibnamefont {Zannoni}},\
  }\href@noop {} {\bibfield  {journal} {\bibinfo  {journal} {Soft Matter},\
  }\textbf {\bibinfo {volume} {4}},\ \bibinfo {pages} {2030} (\bibinfo {year}
  {2008})}\BibitemShut {NoStop}%
\bibitem [{\citenamefont {Sawamura}\ \emph {et~al.}(2002)\citenamefont
  {Sawamura}, \citenamefont {Kawai}, \citenamefont {Matsuo}, \citenamefont
  {Kanie}, \citenamefont {Kato},\ and\ \citenamefont
  {Nakamura}}]{Sawamura2002}%
  \BibitemOpen
  \bibfield  {author} {\bibinfo {author} {\bibfnamefont {M.}~\bibnamefont
  {Sawamura}}, \bibinfo {author} {\bibfnamefont {K.}~\bibnamefont {Kawai}},
  \bibinfo {author} {\bibfnamefont {Y.}~\bibnamefont {Matsuo}}, \bibinfo
  {author} {\bibfnamefont {K.}~\bibnamefont {Kanie}}, \bibinfo {author}
  {\bibfnamefont {T.}~\bibnamefont {Kato}}, \ and\ \bibinfo {author}
  {\bibfnamefont {E.}~\bibnamefont {Nakamura}},\ }\href@noop {} {\bibfield
  {journal} {\bibinfo  {journal} {Nature},\ }\textbf {\bibinfo {volume}
  {419}},\ \bibinfo {pages} {702} (\bibinfo {year} {2002})}\BibitemShut
  {NoStop}%
\bibitem [{\citenamefont {Simpson}\ \emph {et~al.}(2004)\citenamefont
  {Simpson}, \citenamefont {Wu}, \citenamefont {Watson},\ and\ \citenamefont
  {M{\"u}llen}}]{simpson2004}%
  \BibitemOpen
  \bibfield  {author} {\bibinfo {author} {\bibfnamefont {C.~D.}\ \bibnamefont
  {Simpson}}, \bibinfo {author} {\bibfnamefont {J.}~\bibnamefont {Wu}},
  \bibinfo {author} {\bibfnamefont {M.~D.}\ \bibnamefont {Watson}}, \ and\
  \bibinfo {author} {\bibfnamefont {K.}~\bibnamefont {M{\"u}llen}},\
  }\href@noop {} {\bibfield  {journal} {\bibinfo  {journal} {J. Mater. Chem.},\
  }\textbf {\bibinfo {volume} {14}},\ \bibinfo {pages} {494} (\bibinfo {year}
  {2004})}\BibitemShut {NoStop}%
\bibitem [{\citenamefont {Xu}\ and\ \citenamefont {Swager}(1993)}]{xu1993rbl}%
  \BibitemOpen
  \bibfield  {author} {\bibinfo {author} {\bibfnamefont {B.}~\bibnamefont
  {Xu}}\ and\ \bibinfo {author} {\bibfnamefont {T.~M.}\ \bibnamefont
  {Swager}},\ }\href@noop {} {\bibfield  {journal} {\bibinfo  {journal} {J. Am.
  Chem. Soc.},\ }\textbf {\bibinfo {volume} {115}},\ \bibinfo {pages} {1159}
  (\bibinfo {year} {1993})}\BibitemShut {NoStop}%
\bibitem [{\citenamefont {Malthete}\ and\ \citenamefont
  {Collet}(1987)}]{malthete1987icc}%
  \BibitemOpen
  \bibfield  {author} {\bibinfo {author} {\bibfnamefont {J.}~\bibnamefont
  {Malthete}}\ and\ \bibinfo {author} {\bibfnamefont {A.}~\bibnamefont
  {Collet}},\ }\href@noop {} {\bibfield  {journal} {\bibinfo  {journal} {J. Am.
  Chem. Soc.},\ }\textbf {\bibinfo {volume} {109}},\ \bibinfo {pages} {7544}
  (\bibinfo {year} {1987})}\BibitemShut {NoStop}%
\bibitem [{\citenamefont {Rabideau}\ and\ \citenamefont
  {Sygula}(1996)}]{Rabideau1996}%
  \BibitemOpen
  \bibfield  {author} {\bibinfo {author} {\bibfnamefont {P.~W.}\ \bibnamefont
  {Rabideau}}\ and\ \bibinfo {author} {\bibfnamefont {A.}~\bibnamefont
  {Sygula}},\ }\href@noop {} {\bibfield  {journal} {\bibinfo  {journal}
  {Accounts of Chemical Research},\ }\textbf {\bibinfo {volume} {29}},\
  \bibinfo {pages} {235} (\bibinfo {year} {1996})}\BibitemShut {NoStop}%
\bibitem [{\citenamefont {Forkey}\ \emph {et~al.}(1997)\citenamefont {Forkey},
  \citenamefont {Attar}, \citenamefont {Noll}, \citenamefont {Koerner},
  \citenamefont {Olmstead},\ and\ \citenamefont {Balch}}]{Forkey1997}%
  \BibitemOpen
  \bibfield  {author} {\bibinfo {author} {\bibfnamefont {D.~M.}\ \bibnamefont
  {Forkey}}, \bibinfo {author} {\bibfnamefont {S.}~\bibnamefont {Attar}},
  \bibinfo {author} {\bibfnamefont {B.~C.}\ \bibnamefont {Noll}}, \bibinfo
  {author} {\bibfnamefont {R.}~\bibnamefont {Koerner}}, \bibinfo {author}
  {\bibfnamefont {M.~M.}\ \bibnamefont {Olmstead}}, \ and\ \bibinfo {author}
  {\bibfnamefont {A.~L.}\ \bibnamefont {Balch}},\ }\href@noop {} {\bibfield
  {journal} {\bibinfo  {journal} {J. Am. Chem. Soc.},\ }\textbf {\bibinfo
  {volume} {119}},\ \bibinfo {pages} {5766} (\bibinfo {year}
  {1997})}\BibitemShut {NoStop}%
\bibitem [{\citenamefont {Matsuo}\ \emph {et~al.}(2004)\citenamefont {Matsuo},
  \citenamefont {Tahara}, \citenamefont {Sawamura},\ and\ \citenamefont
  {Nakamura}}]{Matsuo2004}%
  \BibitemOpen
  \bibfield  {author} {\bibinfo {author} {\bibfnamefont {Y.}~\bibnamefont
  {Matsuo}}, \bibinfo {author} {\bibfnamefont {K.}~\bibnamefont {Tahara}},
  \bibinfo {author} {\bibfnamefont {M.}~\bibnamefont {Sawamura}}, \ and\
  \bibinfo {author} {\bibfnamefont {E.}~\bibnamefont {Nakamura}},\ }\href@noop
  {} {\bibfield  {journal} {\bibinfo  {journal} {J. Am. Chem. Soc.},\ }\textbf
  {\bibinfo {volume} {126}},\ \bibinfo {pages} {8725} (\bibinfo {year}
  {2004})}\BibitemShut {NoStop}%
\bibitem [{\citenamefont {Sakurai}\ \emph {et~al.}(2005)\citenamefont
  {Sakurai}, \citenamefont {Daiko}, \citenamefont {Sakane}, \citenamefont
  {Amaya},\ and\ \citenamefont {Hirao}}]{Sakurai2005}%
  \BibitemOpen
  \bibfield  {author} {\bibinfo {author} {\bibfnamefont {H.}~\bibnamefont
  {Sakurai}}, \bibinfo {author} {\bibfnamefont {T.}~\bibnamefont {Daiko}},
  \bibinfo {author} {\bibfnamefont {H.}~\bibnamefont {Sakane}}, \bibinfo
  {author} {\bibfnamefont {T.}~\bibnamefont {Amaya}}, \ and\ \bibinfo {author}
  {\bibfnamefont {T.}~\bibnamefont {Hirao}},\ }\href@noop {} {\bibfield
  {journal} {\bibinfo  {journal} {J. Am. Chem. Soc.},\ }\textbf {\bibinfo
  {volume} {127}},\ \bibinfo {pages} {11580} (\bibinfo {year}
  {2005})}\BibitemShut {NoStop}%
\bibitem [{\citenamefont {Kawase}\ and\ \citenamefont
  {Kurata}(2006)}]{Kawase2006}%
  \BibitemOpen
  \bibfield  {author} {\bibinfo {author} {\bibfnamefont {T.}~\bibnamefont
  {Kawase}}\ and\ \bibinfo {author} {\bibfnamefont {H.}~\bibnamefont
  {Kurata}},\ }\href@noop {} {\bibfield  {journal} {\bibinfo  {journal}
  {Chemical Reviews},\ }\textbf {\bibinfo {volume} {106}},\ \bibinfo {pages}
  {5250} (\bibinfo {year} {2006})}\BibitemShut {NoStop}%
\bibitem [{\citenamefont {Cinacchi}\ and\ \citenamefont {van
  Duijneveldt}(2010)}]{Cinacchi2010}%
  \BibitemOpen
  \bibfield  {author} {\bibinfo {author} {\bibfnamefont {G.}~\bibnamefont
  {Cinacchi}}\ and\ \bibinfo {author} {\bibfnamefont {J.~S.}\ \bibnamefont {van
  Duijneveldt}},\ }\href@noop {} {\bibfield  {journal} {\bibinfo  {journal} {J.
  Phys. Chem. Lett.},\ \bibinfo {pages} {787}} (\bibinfo {year}
  {2010})}\BibitemShut {NoStop}%
\bibitem [{\citenamefont {Zoldesi}\ \emph {et~al.}(2006)\citenamefont
  {Zoldesi}, \citenamefont {van Walree},\ and\ \citenamefont {Imhof}}]{Carmen}%
  \BibitemOpen
  \bibfield  {author} {\bibinfo {author} {\bibfnamefont {C.~I.}\ \bibnamefont
  {Zoldesi}}, \bibinfo {author} {\bibfnamefont {C.~A.}\ \bibnamefont {van
  Walree}}, \ and\ \bibinfo {author} {\bibfnamefont {A.}~\bibnamefont
  {Imhof}},\ }\href@noop {} {\bibfield  {journal} {\bibinfo  {journal}
  {Langmuir},\ }\textbf {\bibinfo {volume} {22}},\ \bibinfo {pages} {4343}
  (\bibinfo {year} {2006})}\BibitemShut {NoStop}%
\bibitem [{\citenamefont {Charnay}\ \emph {et~al.}(2003)\citenamefont
  {Charnay}, \citenamefont {Lee}, \citenamefont {Man}, \citenamefont {Moran},
  \citenamefont {Radloff}, \citenamefont {Bradley},\ and\ \citenamefont
  {Halas}}]{Charnay2003}%
  \BibitemOpen
  \bibfield  {author} {\bibinfo {author} {\bibfnamefont {C.}~\bibnamefont
  {Charnay}}, \bibinfo {author} {\bibfnamefont {A.}~\bibnamefont {Lee}},
  \bibinfo {author} {\bibfnamefont {S.-Q.}\ \bibnamefont {Man}}, \bibinfo
  {author} {\bibfnamefont {C.~E.}\ \bibnamefont {Moran}}, \bibinfo {author}
  {\bibfnamefont {C.}~\bibnamefont {Radloff}}, \bibinfo {author} {\bibfnamefont
  {R.~K.}\ \bibnamefont {Bradley}}, \ and\ \bibinfo {author} {\bibfnamefont
  {N.~J.}\ \bibnamefont {Halas}},\ }\href@noop {} {\bibfield  {journal}
  {\bibinfo  {journal} {J. Chem. Phys. B},\ }\textbf {\bibinfo {volume}
  {107}},\ \bibinfo {pages} {7327} (\bibinfo {year} {2003})}\BibitemShut
  {NoStop}%
\bibitem [{\citenamefont {Wang}\ \emph {et~al.}(2004)\citenamefont {Wang},
  \citenamefont {Graugnard}, \citenamefont {King}, \citenamefont {Wang},\ and\
  \citenamefont {Summers}}]{Wang2004}%
  \BibitemOpen
  \bibfield  {author} {\bibinfo {author} {\bibfnamefont {X.~D.}\ \bibnamefont
  {Wang}}, \bibinfo {author} {\bibfnamefont {E.}~\bibnamefont {Graugnard}},
  \bibinfo {author} {\bibfnamefont {J.~S.}\ \bibnamefont {King}}, \bibinfo
  {author} {\bibfnamefont {Z.~L.}\ \bibnamefont {Wang}}, \ and\ \bibinfo
  {author} {\bibfnamefont {C.~J.}\ \bibnamefont {Summers}},\ }\href@noop {}
  {\bibfield  {journal} {\bibinfo  {journal} {nano lett.},\ }\textbf {\bibinfo
  {volume} {4}},\ \bibinfo {pages} {2223} (\bibinfo {year} {2004})}\BibitemShut
  {NoStop}%
\bibitem [{\citenamefont {Liu}\ \emph {et~al.}(2005)\citenamefont {Liu},
  \citenamefont {Maaroof}, \citenamefont {Wieczorek},\ and\ \citenamefont
  {Cortie}}]{Liu2005}%
  \BibitemOpen
  \bibfield  {author} {\bibinfo {author} {\bibfnamefont {J.}~\bibnamefont
  {Liu}}, \bibinfo {author} {\bibfnamefont {A.~I.}\ \bibnamefont {Maaroof}},
  \bibinfo {author} {\bibfnamefont {L.}~\bibnamefont {Wieczorek}}, \ and\
  \bibinfo {author} {\bibfnamefont {M.~B.}\ \bibnamefont {Cortie}},\
  }\href@noop {} {\bibfield  {journal} {\bibinfo  {journal} {Adv. Mater.},\
  }\textbf {\bibinfo {volume} {17}},\ \bibinfo {pages} {1276} (\bibinfo {year}
  {2005})}\BibitemShut {NoStop}%
\bibitem [{\citenamefont {Jagadeesan}\ \emph {et~al.}(2008)\citenamefont
  {Jagadeesan}, \citenamefont {Mansoori}, \citenamefont {Mandal}, \citenamefont
  {Sundaresan},\ and\ \citenamefont {Eswaramoorthy}}]{Jagadeesan2008}%
  \BibitemOpen
  \bibfield  {author} {\bibinfo {author} {\bibfnamefont {D.}~\bibnamefont
  {Jagadeesan}}, \bibinfo {author} {\bibfnamefont {U.}~\bibnamefont
  {Mansoori}}, \bibinfo {author} {\bibfnamefont {P.}~\bibnamefont {Mandal}},
  \bibinfo {author} {\bibfnamefont {A.}~\bibnamefont {Sundaresan}}, \ and\
  \bibinfo {author} {\bibfnamefont {M.}~\bibnamefont {Eswaramoorthy}},\
  }\href@noop {} {\bibfield  {journal} {\bibinfo  {journal} {Angew. Chem. Int.
  Ed.},\ }\textbf {\bibinfo {volume} {47}},\ \bibinfo {pages} {7685} (\bibinfo
  {year} {2008})}\BibitemShut {NoStop}%
\bibitem [{\citenamefont {Love}\ \emph {et~al.}(2002)\citenamefont {Love},
  \citenamefont {Gates}, \citenamefont {Wolfe}, \citenamefont {Paul},\ and\
  \citenamefont {Whitesides}}]{Love2002}%
  \BibitemOpen
  \bibfield  {author} {\bibinfo {author} {\bibfnamefont {J.~C.}\ \bibnamefont
  {Love}}, \bibinfo {author} {\bibfnamefont {B.~D.}\ \bibnamefont {Gates}},
  \bibinfo {author} {\bibfnamefont {D.~B.}\ \bibnamefont {Wolfe}}, \bibinfo
  {author} {\bibfnamefont {K.~E.}\ \bibnamefont {Paul}}, \ and\ \bibinfo
  {author} {\bibfnamefont {G.~M.}\ \bibnamefont {Whitesides}},\ }\href@noop {}
  {\bibfield  {journal} {\bibinfo  {journal} {nano lett.},\ }\textbf {\bibinfo
  {volume} {2}},\ \bibinfo {pages} {891} (\bibinfo {year} {2002})}\BibitemShut
  {NoStop}%
\bibitem [{\citenamefont {Hosein}\ and\ \citenamefont
  {Liddell}(2007)}]{Hosein2007}%
  \BibitemOpen
  \bibfield  {author} {\bibinfo {author} {\bibfnamefont {I.~D.}\ \bibnamefont
  {Hosein}}\ and\ \bibinfo {author} {\bibfnamefont {C.~M.}\ \bibnamefont
  {Liddell}},\ }\href@noop {} {\bibfield  {journal} {\bibinfo  {journal}
  {Langmuir},\ }\textbf {\bibinfo {volume} {23}},\ \bibinfo {pages} {8810}
  (\bibinfo {year} {2007})}\BibitemShut {NoStop}%
\bibitem [{\citenamefont {Higuchi}\ \emph {et~al.}(2006)\citenamefont
  {Higuchi}, \citenamefont {Yabu},\ and\ \citenamefont {Shimomura}}]{higuchi}%
  \BibitemOpen
  \bibfield  {author} {\bibinfo {author} {\bibfnamefont {T.}~\bibnamefont
  {Higuchi}}, \bibinfo {author} {\bibfnamefont {H.}~\bibnamefont {Yabu}}, \
  and\ \bibinfo {author} {\bibfnamefont {M.}~\bibnamefont {Shimomura}},\
  }\href@noop {} {\bibfield  {journal} {\bibinfo  {journal} {Colloids Surf.
  A},\ }\textbf {\bibinfo {volume} {284}},\ \bibinfo {pages} {250} (\bibinfo
  {year} {2006})}\BibitemShut {NoStop}%
\bibitem [{\citenamefont {Lu}\ \emph {et~al.}(2001)\citenamefont {Lu},
  \citenamefont {Yin},\ and\ \citenamefont {Xia}}]{Xia}%
  \BibitemOpen
  \bibfield  {author} {\bibinfo {author} {\bibfnamefont {Y.}~\bibnamefont
  {Lu}}, \bibinfo {author} {\bibfnamefont {Y.}~\bibnamefont {Yin}}, \ and\
  \bibinfo {author} {\bibfnamefont {Y.}~\bibnamefont {Xia}},\ }\href@noop {}
  {\bibfield  {journal} {\bibinfo  {journal} {Adv. Mater.},\ }\textbf {\bibinfo
  {volume} {13}},\ \bibinfo {pages} {34} (\bibinfo {year} {2001})}\BibitemShut
  {NoStop}%
\bibitem [{\citenamefont {Marechal}\ \emph {et~al.}(2010)\citenamefont
  {Marechal}, \citenamefont {Kortschot}, \citenamefont {Demiro{\"o}rs},
  \citenamefont {Imhof},\ and\ \citenamefont {Dijkstra}}]{Marechal2010bowls}%
  \BibitemOpen
  \bibfield  {author} {\bibinfo {author} {\bibfnamefont {M.}~\bibnamefont
  {Marechal}}, \bibinfo {author} {\bibfnamefont {R.~J.}\ \bibnamefont
  {Kortschot}}, \bibinfo {author} {\bibfnamefont {A.~F.}\ \bibnamefont
  {Demiro{\"o}rs}}, \bibinfo {author} {\bibfnamefont {A.}~\bibnamefont
  {Imhof}}, \ and\ \bibinfo {author} {\bibfnamefont {M.}~\bibnamefont
  {Dijkstra}},\ }\href@noop {} {\bibfield  {journal} {\bibinfo  {journal} {nano
  lett.},\ }\textbf {\bibinfo {volume} {10}},\ \bibinfo {pages} {1907}
  (\bibinfo {year} {2010})}\BibitemShut {NoStop}%
\bibitem [{\citenamefont {Widom}(1963)}]{widom}%
  \BibitemOpen
  \bibfield  {author} {\bibinfo {author} {\bibfnamefont {B.}~\bibnamefont
  {Widom}},\ }\href@noop {} {\bibfield  {journal} {\bibinfo  {journal} {J.
  Chem. Phys.},\ }\textbf {\bibinfo {volume} {39}},\ \bibinfo {pages} {2808}
  (\bibinfo {year} {1963})}\BibitemShut {NoStop}%
\bibitem [{\citenamefont {Veerman}\ and\ \citenamefont
  {Frenkel}(1992)}]{Veerman_Frenkel}%
  \BibitemOpen
  \bibfield  {author} {\bibinfo {author} {\bibfnamefont {J.~A.~C.}\
  \bibnamefont {Veerman}}\ and\ \bibinfo {author} {\bibfnamefont
  {D.}~\bibnamefont {Frenkel}},\ }\href@noop {} {\bibfield  {journal} {\bibinfo
   {journal} {Phys. Rev. A},\ }\textbf {\bibinfo {volume} {45}},\ \bibinfo
  {pages} {5632} (\bibinfo {year} {1992})}\BibitemShut {NoStop}%
\bibitem [{\citenamefont {Bates}\ and\ \citenamefont
  {Frenkel}(1998)}]{Bates_Frenkel}%
  \BibitemOpen
  \bibfield  {author} {\bibinfo {author} {\bibfnamefont {M.~A.}\ \bibnamefont
  {Bates}}\ and\ \bibinfo {author} {\bibfnamefont {D.}~\bibnamefont
  {Frenkel}},\ }\href@noop {} {\bibfield  {journal} {\bibinfo  {journal} {Phys.
  Rev. E},\ }\textbf {\bibinfo {volume} {57}},\ \bibinfo {pages} {4824}
  (\bibinfo {year} {1998})}\BibitemShut {NoStop}%
\bibitem [{\citenamefont {Filion}\ \emph {et~al.}(2009)\citenamefont {Filion},
  \citenamefont {Marechal}, \citenamefont {van Oorschot}, \citenamefont {Pelt},
  \citenamefont {Smallenburg},\ and\ \citenamefont
  {Dijkstra}}]{PhysRevLettSSS}%
  \BibitemOpen
  \bibfield  {author} {\bibinfo {author} {\bibfnamefont {L.}~\bibnamefont
  {Filion}}, \bibinfo {author} {\bibfnamefont {M.}~\bibnamefont {Marechal}},
  \bibinfo {author} {\bibfnamefont {B.}~\bibnamefont {van Oorschot}}, \bibinfo
  {author} {\bibfnamefont {D.}~\bibnamefont {Pelt}}, \bibinfo {author}
  {\bibfnamefont {F.}~\bibnamefont {Smallenburg}}, \ and\ \bibinfo {author}
  {\bibfnamefont {M.}~\bibnamefont {Dijkstra}},\ }\href@noop {} {\bibfield
  {journal} {\bibinfo  {journal} {Phys. Rev. Lett.},\ }\textbf {\bibinfo
  {volume} {103}},\ \bibinfo {pages} {188302} (\bibinfo {year}
  {2009})}\BibitemShut {NoStop}%
\bibitem [{\citenamefont {Frenkel}\ and\ \citenamefont
  {Smit}(2002)}]{FrenkelSmit}%
  \BibitemOpen
  \bibfield  {author} {\bibinfo {author} {\bibfnamefont {D.}~\bibnamefont
  {Frenkel}}\ and\ \bibinfo {author} {\bibfnamefont {B.}~\bibnamefont {Smit}},\
  }\href@noop {} {\emph {\bibinfo {title} {{Understanding molecular
  simulation}}}}\ (\bibinfo  {publisher} {Academic Press},\ \bibinfo {year}
  {2002})\BibitemShut {NoStop}%
\bibitem [{\citenamefont {Marechal}\ and\ \citenamefont
  {Dijkstra}(2008)}]{dumbbell_article}%
  \BibitemOpen
  \bibfield  {author} {\bibinfo {author} {\bibfnamefont {M.}~\bibnamefont
  {Marechal}}\ and\ \bibinfo {author} {\bibfnamefont {M.}~\bibnamefont
  {Dijkstra}},\ }\href@noop {} {\bibfield  {journal} {\bibinfo  {journal}
  {Phys. Rev. E},\ }\textbf {\bibinfo {volume} {77}},\ \bibinfo {eid} {061405}
  (\bibinfo {year} {2008})}\BibitemShut {NoStop}%
\bibitem [{\citenamefont {Fortini}\ \emph {et~al.}(2005)\citenamefont
  {Fortini}, \citenamefont {Dijkstra}, \citenamefont {Schmidt},\ and\
  \citenamefont {Wessels}}]{Fortini_soft_pot}%
  \BibitemOpen
  \bibfield  {author} {\bibinfo {author} {\bibfnamefont {A.}~\bibnamefont
  {Fortini}}, \bibinfo {author} {\bibfnamefont {M.}~\bibnamefont {Dijkstra}},
  \bibinfo {author} {\bibfnamefont {M.}~\bibnamefont {Schmidt}}, \ and\
  \bibinfo {author} {\bibfnamefont {P.~P.~F.}\ \bibnamefont {Wessels}},\
  }\href@noop {} {\bibfield  {journal} {\bibinfo  {journal} {Phys. Rev. E},\
  }\textbf {\bibinfo {volume} {71}},\ \bibinfo {pages} {051403} (\bibinfo
  {year} {2005})}\BibitemShut {NoStop}%
\bibitem [{\citenamefont {Parrinello}\ and\ \citenamefont
  {Rahman}(1980)}]{Parrinello}%
  \BibitemOpen
  \bibfield  {author} {\bibinfo {author} {\bibfnamefont {M.}~\bibnamefont
  {Parrinello}}\ and\ \bibinfo {author} {\bibfnamefont {A.}~\bibnamefont
  {Rahman}},\ }\href@noop {} {\bibfield  {journal} {\bibinfo  {journal} {Phys.
  Rev. Lett.},\ }\textbf {\bibinfo {volume} {45}},\ \bibinfo {pages} {1196}
  (\bibinfo {year} {1980})}\BibitemShut {NoStop}%
\bibitem [{\citenamefont {He}\ and\ \citenamefont {Siders}(1990)}]{He1990UFOs}%
  \BibitemOpen
  \bibfield  {author} {\bibinfo {author} {\bibfnamefont {M.}~\bibnamefont
  {He}}\ and\ \bibinfo {author} {\bibfnamefont {P.}~\bibnamefont {Siders}},\
  }\href@noop {} {\bibfield  {journal} {\bibinfo  {journal} {J. Chem. Phys.},\
  }\textbf {\bibinfo {volume} {94}},\ \bibinfo {pages} {7280} (\bibinfo {year}
  {1990})}\BibitemShut {NoStop}%
\bibitem [{\citenamefont {Eppenga}\ and\ \citenamefont
  {Frenkel}(1984)}]{Eppenga}%
  \BibitemOpen
  \bibfield  {author} {\bibinfo {author} {\bibfnamefont {R.}~\bibnamefont
  {Eppenga}}\ and\ \bibinfo {author} {\bibfnamefont {D.}~\bibnamefont
  {Frenkel}},\ }\href@noop {} {\bibfield  {journal} {\bibinfo  {journal} {Mol.
  Phys.},\ }\textbf {\bibinfo {volume} {52}},\ \bibinfo {pages} {1303}
  (\bibinfo {year} {1984})}\BibitemShut {NoStop}%
\end{thebibliography}%

\end{document}